\documentclass[aps, prc, 12pt, nofootinbib, showpacs, superscriptaddress, tightenlines, groupedaddress]{revtex4-2}
\usepackage{amsmath,amssymb,amsbsy,bm}
\usepackage{graphicx}
\usepackage{comment}
\usepackage{float}
\usepackage[colorlinks=true,linkcolor=blue,citecolor=blue,urlcolor=blue]{hyperref}
\usepackage[margin=0.75in]{geometry}
\usepackage{silence}
\begin{document}

\title{The Interplay of Femtoscopic and Charge-Balance Correlations}
\author{Scott Pratt and Karina Martirosova}
\affiliation{Department of Physics and Astronomy and Facility for Rare Isotope Beams\\
Michigan State University, East Lansing, MI 48824~~USA}
\date{\today}


\begin{abstract}
Correlations driven by the constraints of local charge conservation have been shown to provide insight into the chemical evolution and diffusivity of the high-temperature matter created in ultra-relativistic heavy ion collisions. Two-particle correlations driven by final-state interactions have allowed the extraction of critical femtoscopic space-time information about the expansion and dissolution of the same collisions. Whereas correlations from final-state interactions mainly appear at small relative momenta,  a few tens of MeV/$c$, charge-balance correlations extend over a range of hundreds of MeV/$c$. In nearly all previous analyses, this separation of scales is used to focus solely on one class or the other. The purpose of this study is to quantitatively understand the degree to which correlations from final-state interactions distort the interpretation of charge-balance correlations and vice versa.
\end{abstract}

\maketitle

\section{Introduction}\label{sec:intro}
Charge balance correlations are rather simple to understand. For each observed charge, there exists either an additional opposite charge or one fewer charges of the same sign. Because charge is locally conserved, the balancing charge should be found nearby in coordinate space, and because of collective flow, this correlation is mapped onto relative momentum. A charge balance function (BF) binned by relative rapidity and relative azimuthal angle describes the probability of finding the balancing charge at some relative rapidity, $\Delta y$, and relative angle $\Delta\phi$. 
\begin{eqnarray}\label{eq:balancedef}
B(\Delta y,\Delta \phi)&=&\int d\phi_1 dy_1 d\phi_2 dy_2~\delta(y_1-y_2-\Delta y) \delta(\phi_1-\phi_2-\Delta\phi)\\
\nonumber
&&\times\left\{
 \frac{1}{2N_+}\left[P_{-+}(\phi_1,y_1;\phi_2,y_2)-P_{++}(\phi_1,y_1;\phi_2,y_2)\right] \right.\nonumber\\
\nonumber
&&\left. +\frac{1}{2N_-}\left[P_{+-}(\phi_1,y_1;\phi_2,y_2)-P_{--}(\phi_1,y_1;\phi_2,y_2)\right]\right\},\\
\nonumber
N_{\pm}&=&\int d\phi dy~P_\pm(\phi,y).
\end{eqnarray}
The like-sign subtraction effectively identifies the location of the balancing charge on a statistical basis. Thus, $B(\Delta y,\Delta \phi)$ represents the conditional probability density for finding a balancing charge (either an opposite charge or the lowered chance of observing a charge of the same sign) separated by $\Delta\phi$ and $\Delta y$ given the observation of a charge somewhere in the detector. BFs can also be indexed by hadron species, $B_{h|h'}(\Delta y,\Delta\phi)$. This then describes the probability of first observing a hadron species $h'$ or $\bar{h}'$, then finding a particle of opposite charge of species $h$ or $\bar{h}$. Due to the experimental difficulty in identifying hadrons which have decayed, the choice of $h$ and $h'$ is often confined to pions, kaons, protons, and their antiparticles.

Even if two balancing charges are emitted from nearly the same point in coordinate space, they will separate in momentum space due to thermal motion. This separation in momentum space would be of the order of a few hundred MeV/$c$, or equivalently $\lesssim 0.5$ radians or units of rapidity. If the balancing charges were created early and had the opportunity to diffuse far from one another in coordinate space, their final separations in momentum space might extend to twice that amount. The mean width of the BF, $\langle\Delta\phi\rangle$ or $\langle\Delta y\rangle$, i.e. the average separation of balancing charges, provides insight into the diffusivity or of the chemical evolution. Because $\langle\Delta y\rangle$ is relatively more sensitive to whether the particles were created early than $\langle\Delta\phi\rangle$ \cite{Bass:2000az}, one can constrain both the chemistry and diffusivity by BFs by analyzing BFs in terms of both $\Delta y$ and $\Delta\phi$ \cite{Pratt:2015jsa,Pratt:2012dz,Pratt:2019pnd}. This sensitivity is amplified by considering BFs indexed by hadronic species. For example, because strangeness is largely produced early in the collisions, kaon BFs, $B_{K|K}(\Delta\phi,\Delta y)$, are especially useful for extracting the diffusivity \cite{Pratt:2019pnd}. Numerous varieties of BFs have now been measured in heavy-ion collisions both by the STAR Collaboration at the Relativistic Heavy Ion Collider (RHIC) \cite{Westfall:2004jh,STAR:2015ryu,Wang:2011za,Wang:2011bea,Li:2011zzx,STAR:2010plm,Westfall:2004cq,STAR:2003kbb,Wang:2012jua,Abelev:2010ab,Adams:2003kg,Aggarwal:2010ya,Adamczyk:2015yga,STAR:2011ab} and by the ALICE Collaboration at the Large Hadron Collider (LHC) \cite{ALICE:2021hjb,Pan:2018dsq,Alam:2017iom,Weber:2013fla,ALICE:2013vrb,Weber:2012ut,JinjinPanThesis,Abelev:2013csa}. At lower energies, the NA49 Collaboration has also measured BFs at CERN SPS energies \cite{Alt:2004gx,Alt:2007hk}. Detailed theoretical models describing the dynamics of charge correlations, superimposed onto state-of-the-art dynamical descriptions of the bulk evolution, have been able to quantitatively reproduce several features of measurements at both RHIC and the LHC \cite{Pratt:2017oyf,Pratt:2018ebf,Pratt:2019pnd,Pratt:2021xvg}. The inferred diffusivity and chemistry from comparing models to data appears consistent with expectations from lattice gauge theory \cite{Borsanyi:2011sw,Aarts:2014nba,Amato:2013naa}.

Correlations at small relative momentum are dominated by the effects of final-state interactions (FSI). The correlations provide detailed spatial and geometric information describing the emission of final-state hadrons. Analyses of this class of correlations is often referred to as femtoscopy. Femtoscopic correlations are typically predicted through the Koonin formula \cite{Koonin:1977fh,Lisa:2005dd},
\begin{eqnarray}\label{eq:koonin}
C_{ab}(\vec{p}_a,\vec{p}_b)&\equiv&\frac{N_{ab}(\vec{p}_a,\vec{p}_b)}{N_a(\vec{p}_a)N_b(\vec{p}_b)},\\
\nonumber
&\approx&\int d^3r~S_{ab}(\vec{p}_a,\vec{p}_b,\vec{r})\left|\phi(\vec{q}',\vec{r})\right|^2,\\
\nonumber
S_{ab}(\vec{p}_a,\vec{p}_b,\vec{r})&=&\frac{\int d^4r_ad^4r_b~s_a(r_a,\vec{p}_a)s_{b}(r_b,\vec{p}_b)
\delta(\vec{r}-\vec{r}'_a+\vec{r}'_b)}{\int d^4r_ad^4r_b~s_a(r_a,\vec{p}'_a)s_{b}(r_b,\vec{p}_b)}.
\end{eqnarray}
Here, $s_a(r,\vec{p})$ describes the probability of emitting a hadron of type $a$ from space-time point $r$ with momentum $\vec{p}$, and $\phi(\vec{q},\Delta\vec{r})$ is the outgoing wave function for two particles with relative separation $\vec{r}$ and relative momentum $\vec{q}=(\vec{p}_1-\vec{p}_2)/2$, as measured in the center-of-mass of the pair frame. The primed coordinates represent the positions of emission in that frame. The function $S_{ab}(\vec{r})$ represents the probability that two particles, one of type $a$ and one of type $b$, are emitted at points separated by $\vec{r}$ in the two-particle center-of-mass frame. It is often referred to as the ``source function'', though that is a misnomer because it does not represent the probability density of the emission function. More accurately, if you assume $\vec{q}'$ is small, its dependence on $\vec{r}$ represents the probability that two particles moving with the same velocity in the asymptotic state would be separated by $\vec{r}$. Generally, the goal of femtoscopic analyses is to extract information about $S_{ab}(\vec{r})$ from measurements of $C_{ab}(\vec{q})$.

There are variants of this formula, but they tend to all become equal in the limit that $\vec{q}$ is small \cite{Pratt:1997pw}. The correlation would be unity if the relative wave function were that of a plane wave, but due to final-state interactions and symmetrization of the outgoing wave function,  $|\phi(\vec{q}),\vec{r})|^2$ differs from unity and provides a correlation which is stronger when the relative positions, $\vec{r}$, are smaller.  Thus, one gains insight into the spatial extent of $s(r,\vec{p})$. For identical pions, the wave function is symmetrized. If one neglects the Coulomb and strong interaction between the pions the squared wave function is then
\begin{eqnarray}
|\phi(\vec{q},\vec{r})|^2&=1+\cos(2\vec{q}\cdot\vec{r}),
\end{eqnarray}
and one can Fourier transform the correlation function to determine the source function, as long as one assumes that there is little dependence of $S_{ab}$ on $q$. One could then extract both size and shape information about $S_{ab}(\vec{r})$.

The space-time characteristics of the emission provide insight into the equation of state \cite{Pratt:1986cc,Pratt:2015zsa}. For example, if the equation of state is soft, the expansion is slow and there is an evaporative nature to the emission. In that case, two pions of identical velocity, $\vec{v}$, might be separated by a large distance due to one pion being emitted early and the other coming late. The spatial separation is large along the direction of $\vec{v}$, while being more compact in the other directions. A more explosive source would result in a more compact spread. Further, for pions with higher velocity, compared to the expansion velocity, emission is increasingly confined to the surface of the expanding fireball. This results in source sizes that fall with increasing transverse momentum \cite{Pratt:1984su}.

Femtoscopic correlations are driven by three types of FSI: symmetrization or anti-symmetrization of wave functions of identical particles, strong interaction, and Coulomb repulsion. Symmetrization effects extend out to relative momenta of $1/R$, where $R\sim 5$ fm is a typical characteristic size. This contribution to the correlation largely vanishes for $|\vec{q}|\gtrsim 50$ MeV/$c$. Strong interactions at low relative momentum are especially important because of the reduced phase space of the background. The two-proton correlation function has a peak at $q\sim 22$ MeV/$c$. The effect of strong interactions at higher relative momentum tends only to appear for well defined resonances, but those resonances, unlike the $pp$ peak at $q\sim 22$ MeV/$c$, are typically included in BF analyses. The third class of FSI derives from the Coulomb interaction. For the Coulomb interaction the correlation extends to larger relative momentum, because the squared wave function behaves as $1\pm me^2/q^2R$ at larger $q$. This is small due to the factor $e^2$, but it becomes the dominant source of FSI at large $q$. Compared to correlation functions, BFs have an extra factor describing the background probability of observing a particle in the bin. If binning by  the magnitude of the relative momentum, $Q_{\rm inv}$, this factor grows quadratically with $Q_{\rm inv}$ due to phase space. Thus, compared to correlations functions, at least visually, BFs tend to magnify the strength of the Coulomb tail. Thus, special care must be given to Coulomb correlations when considering the effects of FSI on BFs. This includes accounting for the fact that any charged particle is accompanied by an oppositely charged balancing particle, which effectively screens the Coulomb interaction with third bodies \cite{Pratt:2003gh}.

A fourth type of interaction affects both charge-balance and femtoscopic correlations. That is annihilation. Within the context of a BF, annihilation is simply a negative source for pair creation. This leads to a dip in the BF at small relative momentum. However, if the annihilation involves particles, e.g. a proton and an anti-proton, that would not have interacted with other particles had they annihilated, the annihilation might have been considered as part of the final-state interaction wave function. For example, Eq. (\ref{eq:koonin}) could apply a relative wave function calculated from a complex potential \cite{Zbroszczyk:2011jv,Kisiel:2004fcn}, with the imaginary part of the potential accounting for the annihilation. The separation of annihilation into a final-state effect, which can be treated quantum mechanically, vs. a negative source function can be blurry. The issue is further complicated by the fact that particles can be regenerated. i.e., if a baryon and anti-baryon can decay to five hadrons, five hadrons can combine to form a baryon anti-baryon pair \cite{Rapp:2001bb,Pan:2014caa,Steinheimer:2017vju,Savchuk:2021aog}. At chemical equilibrium, the rate and the inverse rates are equal. But at final breakup, chemical equilibrium no longer holds and annihilation is more prevalent.

Femtoscopic correlations are constructed to be dimensionless quantities, whereas BFs have units of density per unit rapidity, relative angle or relative momentum. This comes from the fact that Eq. (\ref{eq:koonin}) has two powers of $N_h(\vec{p})$ in the denominator whereas the definition of BFs in Eq. (\ref{eq:balancedef}) has one power. The next section describes how charge-balance correlations and femtoscopic correlations are related.

The basic theory of charge balance calculations is reviewed in Sec. \ref{sec:theory_bf} while Sec. \ref{sec:algorithm_bf}  presents algorithm for calculating BFs from a blastwave calculation.  The theory and method for accounting for the screening of Coulomb interactions is presented in Sec. \ref{sec:theory_screening}. Results of calculations illustrating how correlations from FSI distort BFs are given in Sec. \ref{sec:results_femtodistortions} while Sec. \ref{sec:results_bfdistortions} shows how femtoscopic correlations at small relative momentum are affected by charge balance correlations. A summary, Sec. \ref{sec:summary}, is followed by two appendices, reviewing classical Coulomb correlations and the blast-wave fitting procedure respectively.


\section{Relating Charge-Balance and Femtoscopic Correlations}\label{sec:theory1}

Femtoscopic correlations are nearly always analyzed as a function of relative momentum. Typically, the range of relative momentum under consideration is $0< q\lesssim 100$ MeV/$c$. By focusing on small relative momentum, one can better justify the approximation that the particles interact mainly with one another between the last interaction and the detector. In contrast, BFs are usually analyzed as a function of relative rapidity or relative azimuthal angle. They are sometimes binned by relative momentum, in which case the range of relative momenta tends to be in the range of several hundreds of MeV/$c$, which is the range of the thermal smearing of the space-time correlations. 

Whereas femtoscopic correlations are constructed by dividing the two-particle distribution by an uncorrelated two-particle distribution, BFs are created by dividing by one single-particle distribution. Thus, BFs can be thought of as a ``conditional distribution'', i.e. given the observation of a charge, what is the probability of finding more charges of the opposite sign than of the same sign as a function of relative rapidity or relative azimuthal angle. The two forms are related by factors of the multiplicity,
\begin{eqnarray}
B(p_1|p_2)&=&\frac{1}{2}C_{+-}(p_1,p_2)\frac{dN_+}{dp_1}
+\frac{1}{2}C_{-+}(p_1,p_2)\frac{dN_-}{dp_1}
-\frac{1}{2}C_{++}(p_1,p_2)\frac{dN_+}{dp_1}
-\frac{1}{2}C_{--}(p_1,p_2)\frac{dN_-}{dp_1}.
\end{eqnarray}
The variables $p_1$ and $p_2$ could be any measure of the momentum. As an example, to get BFs binned by relative rapidity, $p_2$ might refer to any momentum in the detector and $p_1$ could refer to the relative rapidity. The quantity $dN_\pm/dp_1$ would then represent the number of charges of type $\pm$ that would have the desired relative rapidity in a single event in the absence of correlation.

Similarly, one can generate correlation functions from BFs, but only the differences between same-sign and opposite sign correlations, and only for the case that the correlations are unchanged if positive and negative particles are switched, i.e. $C_{+-}=C_{-+}$ and $C_{++}=C_{--}$. In that case
\begin{eqnarray}
C_{\rm opp.~sign}(p_1,p_2)-C_{\rm same~sign}(p_1,p_2)&=&
\frac{2B(p_1|p_2)}{dN_-/dp_1+dN_+/dp_1}.
\end{eqnarray}
For a cylindrically symmetric boost-invariant distribution, which will be assumed throughout this paper, one can derive simple relations when the BF is binned by $\Delta y,\Delta\phi$ or $Q_{\rm inv}$,
\begin{eqnarray}\label{eq:RBC}
B_{h|h'}(\Delta y)&=&\left(\frac{dN_{h+}}{dy}+\frac{dN_{h-}}{dy}\right)
\left[C_{h,h',{\rm opp.sign}}(\Delta y)+C_{h,h',{\rm same.sign}}(\Delta y)\right],\\
\nonumber
B_{h|h'}(\Delta \phi)&=&\frac{2Y_{\rm max}}{\pi}
\left(\frac{dN_{h+}}{dy}+\frac{dN_{h-}}{dy}\right)
\left[C_{h,h',{\rm opp.sign}}(\Delta \phi)+C_{h,h',{\rm same.sign}}(\Delta \phi)\right],\\
\nonumber
B_{h|h'}(Q_{\rm inv})&=&2Y_{\rm max}P_{h,h'}(Q_{\rm inv})
\left(\frac{dN_{h+}}{dy}+\frac{dN_{h-}}{dy}\right)
\left[C_{h,h',{\rm opp.sign}}(Q_{\rm inv})+C_{h,h',{\rm same.sign}}(Q_{\rm inv})\right].
\end{eqnarray}
Here, $Y_{\rm max}$ is the range of the acceptance in rapidity, $-Y_{\rm max}<y<Y_{\rm max}$. The expressions are built on the assumption that the correlation functions are corrected for the acceptance in rapidity, meaning that for any charge with rapidity $y_1$, all charges with rapidities $y_2$, within the range of $y_1-2Y_{\rm max}<y_2<y_1+2Y_{\rm max}$ are assumed to have been measured. The dependence on $Y_{\rm max}$ is especially important for $B_{h|h'}(\Delta \phi)$. If one were to increase the range in rapidity the correlations binned by $\Delta\phi$ would be diluted as would the charge balance functions. As long as correlations in $Q_{\rm inv}$ and $\Delta y$ do not extend beyond $2Y_{\rm max}$, $B(Q_{\rm inv})$ and $B(\Delta y)$ are largely independent of $Y_{\rm max}$. It should be emphasized that $\Delta y$ and $\Delta \phi$ refer to the absolute values of relative rapidity and relative azimuthal angle. Otherwise, the first two expressions in Eq. (\ref{eq:RBC}) would include an extra factor of $1/2$, and the remaining half the strength would be found at negative values of $\Delta y$ and $\Delta \phi$. The relative momentum $Q_{\rm inv}$ is the magnitude of the relative momentum, $|\vec{p}-\vec{p}'|$, in the frame of the pair. Finally, $P_{h,h'}(Q_{\rm inv})$ refers to the probability density of any two particles being separated by $Q_{\rm inv}$, where the second particle is randomly boosted so that its rapidity is uniformly found in rapidity acceptance as described above. Because $P_{h,h'}$ will fall inversely with $Y_{\rm max}$, the product $Y_{\rm max}P_{h,h'}(Q_{\rm inv})$ in the expression for $B_{h|h'}(Q_{\rm inv})$ in Eq. (\ref{eq:RBC}) is independent of $Y_{\rm max}$ once $Y_{\rm max}$ is large enough to capture all the pairs for the specific $Q_{\rm inv}$.

\section{Review of Charge Balance Correlations}\label{sec:theory_bf}

A hadron of type $h$ with charge $q_{hu}, q_{hd}$ and $q_{hs}$ (where $u,d,s$ refers to the up, down and strange charges) must be accompanied by balancing charges, carried by the altered distributions of other hadrons. Here, we show what number of hadrons $\delta N_{h'}$ result from the existence of $\delta N_h$. This is represented by $\kappa_{h'|h}$ where
\begin{eqnarray}
\delta N_{h'}&=&\kappa_{h'|h}\delta N_h.
\end{eqnarray}
First, we express $\delta N_{h'}$ in terms of a small chemical potential for a thermal system. We find the three chemical potentials, $\mu_u,\mu_d$ and $\mu_s$, necessary to produce the correct amount of balancing charge.
The number of hadrons of species $h'$ is altered by the presence of a hadron $h$ according to
\begin{eqnarray}
\delta N_{h'}&=&\langle N_h\rangle\frac{\mu_aq_{h'a}}{T},
\end{eqnarray}
where $q_{ha}$ is the charge of type $a$ on a hadron of type $h$, with $a=u,d$ or $s$. This is a thermal argument where the number of hadrons of a species $h'$ is altered by a factor $e^{\mu_aq_{h'a}/T}\approx 1+\mu_aq_{h'a}/T$.

Summing over all the charges from all the hadrons $h'$ should yield the charge that balances that carried by $h$,
\begin{eqnarray}
-q_{ha}&=&\sum_{h'}\delta N_{h'}q_{h'a}\\
\nonumber
&=&\sum_{h'}\langle N_{h'}\rangle\frac{\mu_b}{T}q_{h'b}q_{h'a}\\
\nonumber
&=&V\chi_{ab}\frac{\mu_b}{T},\\
\nonumber
\frac{\mu_a}{T}&=&-\frac{1}{V}\chi^{-1}_{ab}q_{hb}.
\end{eqnarray}
Here, the charge susceptibility of a non-interacting gas is
\begin{eqnarray}\label{eq:chihadrongas}
\chi_{ab}&=&\frac{1}{V}\langle Q_aQ_b\rangle=\sum_h \langle n_h\rangle q_{ha}q_{hb},
\end{eqnarray}
or equivalently, the charge correlation for a non-correlated hadron gas is confined to charges within the same hadron. This then provides the altered number of hadrons of type $h'$ due to the existence of a single hadron of type $h$,
\begin{eqnarray}
\delta N_{h'}&=&-\sum_{ab}\chi^{(-1)}_{ab}\langle n_{h'}\rangle q_{h'a}q_{hb},\\
\nonumber
&=&\kappa_{h'|h}\delta N_h,\\
\nonumber
\kappa_{h'|h}&=&-\langle n_{h'}\rangle\sum_{ab}\chi^{-1}_{ab}q_{h'a}q_{hb}.
\end{eqnarray}
Here, $\langle n_h\rangle$ is the mean density of hadrons of type $h$. The kernel $\kappa_{h'|h}=-\kappa_{h'|\bar{h}}$ because $h$ and its antiparticle $\bar{h}$ have opposite charges. Thus, for any BF,
\begin{eqnarray}\label{eq:kappa}
\int dp'~B_{h'|h}(p'|p)&=&  \frac{1}{4}\left[\kappa_{h'|h}+\kappa_{\bar{h}'|\bar{h}}-\kappa_{h'|\bar{h}}-\kappa_{\bar{h}'|h}\right].
\end{eqnarray}
One can quickly check to see that if one were to sum the normalizations over all $h'$ multiplied by $q_{h'a}$, one would indeed find the charge $q_{ha}$,
\begin{eqnarray}
\sum_{h'}q_{h'a}\int dp'~B_{h'|h}(p'|p)&=&
\frac{1}{4}\sum_{h'}q_{h'a}\left[\kappa_{h'|h}+\kappa_{\bar{h}'|\bar{h}}-\kappa_{h'|\bar{h}}-\kappa_{\bar{h}'|h}\right]\\
\nonumber
&=&\sum_{h'}\sum_{h}q_{h'a}\langle n_{h'}\rangle q_{h'a'}\chi^{-1}_{a'b}q_{hb}\\
\nonumber
&=&\chi_{aa'}\chi^{-1}_{a'b}q_{hb}\\
\nonumber
&=&q_{ha}.
\end{eqnarray}

The expressions above ignore decays. Decays can be included by altering the kernels $\kappa_{h'|h}$ to include both the contribution where $h'$ and $h$ come from the same decaying parent, and the case where two charges correlated as described by the kernel $\kappa$ above then decay to $h'$ and $h$. If a hadron $H$ decays into a set of channels $c_H$, where each channel has a branching ratio $b_{c_H}$, and if the number of hadrons of type $h$ coming from the particular channel is $m_{c_H}$, the contribution to the  kernel $K(h'|h)$ from decays is
\begin{eqnarray}
K^{\rm(d)}_{h'|h}&=&\frac{1}{\langle\langle N_h\rangle\rangle}\sum_H \langle N_H\rangle b_{c_H}m_{c_H,h}m_{c_H,h'}~,\\
\nonumber
\langle\langle N_h\rangle\rangle&=&\sum_{H,c_H}\langle N_H\rangle m_{c_H,h}b_{c_H}.
\end{eqnarray}
The channels $c_H$ include the case where a particle is stable, i.e. where $H=h$ and there are no additional products. The notation $\langle\langle N_h\rangle\rangle$ signifies that this is the multiplicity after decays have taken place, whereas $\langle N_h\rangle$ signifies the density at the time hadrons were created with balancing charge assigned according to the arguments above.

One can then add in the contribution from correlations from charge balance at hadronization to find the complete kernel, $K$,
\begin{eqnarray}
K_{h'|h}&=&K^{\rm(d)}_{h'|h}\\
\nonumber
&+&\frac{1}{\langle\langle N_h\rangle\rangle}\sum_{H,c_H,H',c_{H'}} \kappa_{H'|H} \langle N_H\rangle
b_{c_H} m_{c_H,h} \langle N_{H'}\rangle b_{c_{H'}} m_{c_{H'},h'}.
\end{eqnarray}
The normalization of the BF with decays included is 
\begin{eqnarray}\label{eq:KZ}
Z_{h'|h}&=&\frac{1}{4}\left\{
K_{h'|h}-K_{h'|\bar{h}}+K_{\bar{h}'|\bar{h}}-K_{\bar{h}'|h}.
\right\}.
\end{eqnarray}

For use later on, a function is defined that is symmetric in $h$ and $h'$,
\begin{eqnarray}\label{eq:KW}
W_{h'|h}&\equiv&\frac{\langle\langle N_h\rangle\rangle}{\langle\langle N\rangle\rangle}K_{h'|h}\label{eq:Weightdef}\\
\nonumber
&=&\left\{\sum_{H,c_H} \frac{\langle n_H\rangle}{\langle n\rangle} b_{c_H}m_{c_H,h} m_{c_H,h'}
+\sum_{H,c_H,H',c_{H'}}\frac{\langle n_H\rangle\langle n_{H'}\rangle}{\langle n\rangle^2}w_{H',H}b_{c_H}m_{c_H,h}b_{c_{H'}}m_{c_{H'},h'}\right\},\\
\nonumber 
w_{H',H}&=&
\left[\langle n\rangle(q_{Ha}\chi^{-1}_{ab}q_{H'b})\right].
\end{eqnarray}
Here, $\langle n_H\rangle$ is the density of hadrons of species $H$ and $\langle n\rangle$ is the net hadron density, at the time chemical equilibrium is lost. $\langle\langle N_h\rangle\rangle/\langle\langle N\rangle\rangle$ is the ratio of hadrons of type $h$ to total hadrons in the final state,
\begin{eqnarray}
\langle\langle N\rangle\rangle&=&\sum_h\langle\langle N_h\rangle\rangle.
\end{eqnarray}

\section{Calculating BFs from Blast Wave Model}\label{sec:algorithm_bf}

By inspection of the expression for $W_{h'|h}$ in Eq. (\ref{eq:Weightdef}) and the way in which it relates to $K_{h'|h}$ in Eq. (\ref{eq:KW}), one can see that the BFs can be generated by a two step process. In the first step the contribution from decays, the first sum in Eq. (\ref{eq:Weightdef}), is calculated. For this case one generates initial hadrons $H$ according to the weight $\langle n_h\rangle/\langle n\rangle$. This is simply choosing the resonance species proportional to the density. One then decays the resonance into a given branch according to $d(H,\langle h_1\cdots h_m\rangle)$. Summing through the products one increments the BF according to the relative momentum. This results in a representation of the first term of $W_{h'|h}$ binned by relative momentum for each species combination. 

To generate the second term, one generates two particles, of species $H$ and $H'$, randomly according to $\langle n_h\rangle/\langle n\rangle$. Both are then decayed and two sets of products are found. For each combination of a hadron derived from $H$ and a second hadron derived from $H'$, one increments the BF. But, in this case there is an additional weight, $w_{HH'}$ defined in Eq. (\ref{eq:KW}). The bin for the BF is then incremented by $w_{HH'}$. Finally, because this procedure generates a two-particle distribution representing $W_{h'|h}$ instead of $K_{h'|h}$, the distribution is multiplied by the factor $\langle\langle N\rangle\rangle/\langle\langle N_h\rangle\rangle$ as expressed in Eq. (\ref{eq:KW}).

In order to calculate BFs, one needs a model to generate how balancing particles are correlated in momentum space. For the studies here, a simple blast wave prescription is used. Blast wave prescriptions are based on a parametric description of a hot expanding gas. The particular description used here is described further in Appendix \ref{app:blastwave}. Blast wave pictures all have at least three parameters. The first is $T_c$, the temperature which describes the chemical make-up of the emitted hadrons. This is often referred to as the chemical freeze-out temperature. A second parameter denotes the temperature at which the system uncouples, $T_f$, usually referred to as the kinetic freeze-out temperature. This temperature only affects the momenta at which particles are emitted in the frame of the fluid from which they emerge. The third parameter describes the amount of collective transverse flow, $U_\perp$. Higher values of $U_\perp$ or $T_f$ increase the mean momenta and energy of the particles. The system also has longitudinal flow, which is assumed to be of a boost-invariant nature according to a Bjorken prescription, \cite{Bjorken:1982qr}. Two other parameters, the transverse size in coordinate space $R$ and the breakup time $\tau_f$ affect the femtoscopic measurements. Particles are generated in pairs. The mean transverse positions of the pair is chosen according to a uniform distribution with radius $R$, and the collective velocities at those points are assigned the values,
\begin{eqnarray}
u_x&=&U_\perp\frac{x}{R},~u_y=U_\perp\frac{y}{R}.
\end{eqnarray}
The particles are then assigned a position close to the mean position, being adjusted by a random step, $\delta x$ and $\delta y$, where the steps are randomly chosen according to a Gaussian distribution, $P(\delta x)\sim e^{-\delta x^2/2\sigma_R}$, and similarly for $\delta y$. The parameter $\sigma_R$ describes the range over which charge conservation is enforced in the transverse plane. Due to boost invariance, the mean spatial rapidity of the pair can be set to zero, with the particles being placed nearby according to steps $\delta\eta$, which are randomly assigned proportional to $e^{-\delta\eta^2/2\sigma_\eta^2}$. Certainly, more sophisticated blast-wave prescriptions exist, some even taking into account anisotropies in elliptic shape and flow \cite{Retiere:2003kf}. However, the main purpose here is to determine whether femtoscopic correlations sufficiently affect BF observables, so any prescription which approximately reproduces the femtoscopic correlations should suffice.

BFs can be generated for specific species, $B_{h'|h}$, and can be generated by the method enumerated below. Here, the species $h'$ and $h$ are typically chosen to be of opposite sign, so that the BF is positive represents the enhancement for finding an opposite charge. Examples are $\pi^+\pi^-, K^+K^-, p\bar{p}, \pi^+K^-, \pi^+\bar{p}$ and $K^+\bar{p}$. 

\begin{enumerate}
\item Beginning with a list of hadrons, their masses, degeneracies and charges, one calculates the charge susceptibility matrix at the temperature $T_c$, the last temperature for which chemical equilibrium was maintained. This involved calculating the density of each species, then using Eq. (\ref{eq:chihadrongas}) which provides the susceptibility, or charge fluctuation, for an non-interacting hadron gas. Using that susceptibility, one calculates $W_{HH'}$ according to Eq. (\ref{eq:Weightdef}).

\item The contribution to the BF from decays, the first term in Eq. (\ref{eq:Weightdef}), is calculated using Monte Carlo sampling. A number of initial hadrons, $N_{mc}$, are generated. The species is chosen proportional to $\langle n_h\rangle/\langle n\rangle$ which is calculated at temperature $T_c=150$ MeV. All decay products with lifetimes less than 100 fm/$c$ are then simulated. The particles are then randomly placed in coordinate space at position $r$ according to the Gaussian size, $R$. This then gives the transverse velocity as the transverse collective flow increases as $u_\perp=U_\perp r/R$, where $U_\perp$ is a model parameter. Using the freeze-out temperature $T_f$ and $u_\perp$, the momentum is generated. The final decays, long-lived decays are then generated. If the two species $h$ and $h'$ both appear in the final products an array is incremented. The array represents the function $W_{h'|h}$ in Eq. (\ref{eq:Weightdef}), but is binned by whatever kinematic variable is being considered, e.g. relative rapidity. One also increments a counter of $N_h$ and $N_{h'}$. After sufficient sampling, the binning of $W_{h'|h}$ is translated into a binning of $Z_{h'|h}$ according to Eq.s (\ref{eq:KZ}) and (\ref{eq:KW}), which involves dividing by a factor $\langle\langle N_h\rangle\rangle/\langle\langle N\rangle\rangle$. The array is also divided by $N_{mc}$. This then provides the contribution to $B_{h'|h}$ from decays. 

\item The contribution to $B_{h'|h}$ from the second term in Eq. (\ref{eq:Weightdef}) is then calculated. This also is calculated in a Monte Carlo procedure. First, two particles are generated independently, with species $H$ and $H'$. They are chosen according to the thermal weights consistent with the temperature $T_c$. Decay products for each particle are then chosen according to the branching ratios. Again, at this point decays with lifetimes greater than 100 fm/$c$ are not performed. Transverse spatial coordinates are then chosen according to a Gaussian radius, $(R^2-\sigma_R^2)^{-1/2}$. The spatial rapidity is then set to zero. The decay products of $H$ are then all positioned at a position which differs from the original position by a random Gaussian step set by $\sigma_R$ transversely and $\sigma_\eta$ longitudinally. The descendents of $H'$ are placed at  different points. These points are based off the same original position, but with different random steps. The array representing $W_{h'|h}$ is then incremented, but the array elements are not incremented by unity, but instead by $W_{HH'}$. Again, the array for $N_h$ is incremented. Finally, the array is divided by $N_{mc}$ and the factor $\langle\langle N_h\rangle\rangle/\langle\langle N\rangle\rangle$. Finite acceptance is only crudely taken into account by ignoring any pairs with relative rapidity greater than $2Y_{\rm max}$, with $Y_{\rm max}=0.9$, corresponding to the STAR acceptance. This ignores the $p_t$ dependence of the acceptance and efficiency. Even if experiments were to correct for acceptance and efficiency, this calculation would be questionable due to the fact that low $p_t$ particles are not measured and because the low $p_t$ cutoff depends strongly on rapidity, especially for heavier particles. 
\end{enumerate}

\begin{eqnarray}
P_h&=&\frac{\langle\langle N_h\rangle\rangle}{\langle\langle N\rangle\rangle}.
\end{eqnarray}
The two momenta would be chosen to be correlated in coordinate space according to some prescription or model. In this study a blast-wave model is used to represent the probability of emitting particles from a given point with a given momentum. The pair is correlated in coordinate space according to a Gaussian form. The model is described in the Appendix \ref{app:blastwave}. After generating the pair, one would increment a bin of some distribution according to some kinematic variable, e.g. relative rapidity. Instead of incrementing the bin by unity one could increment the bin by the weight $W$ defined in Eq. (\ref{eq:Weightdef}).

Much more realistic models of BFs have been construced and analyzed, e.g. \cite{Pratt:2018ebf,Pratt:2019pnd,Pratt:2021xvg}. These more sophisticated treatments  account for the difference between the distance scales over which strangeness, electric charge or baryon number are conserved. Decays are more realistically taken into account and experimental acceptance and efficiency are considered in detail. More sophisticated treatments can lead to BF widths changing by a few tens of percent from the models used here. But the much simpler, much less numerically intensive, model used here is sufficient to satisfy the goal of this study, which is to understand the degree to which FSI and BF correlations must be simultaneously considered. Comparison with experimental data is not the goal of this study.


\section{Screening Final-State Coulomb Interactions} \label{sec:theory_screening}

Correlations from FSI can be calculated according to a number of methods, which tend to become equal when the relative momentum is small \cite{Pratt:1997pw}. For larger relative momenta, the main method is to generate a pair of hadrons, independent of one another, with momentum $\vec{p}_1$ and $\vec{p}_2$, from space-time coordinates $x_1$ and $x_2$. One then increments a two-particle distribution by an amount $|\phi_{hh'}(\vec{q},\vec{r})|^2$. The distribution is typically binned by relative momentum, but could be binned by some other variable such as relative rapidity. Here, $\vec{q}$ and $\vec{r}$ refer to the relative momentum and position in the center-of-mass of the pair. Because $q\ne 0$, the relative position depends on the time at which $\vec{r}$ is calculated. Here, it is assigned the value corresponding to the separation of the two trajectories at a time half way between the two emissions, and $\vec{q}$ and $\vec{r}$ are calculated in the pair's rest frame. The correlation function is then the average of $|\phi|^2$ within each bin. This method provides a realization of Eq. (\ref{eq:koonin}). 

The squared wave functions differ from unity due to the symmetrization, or anti-symmetrization, of the wave functions, the strong interaction, and the Coulomb force between the two particles. Symmetrization and anti-symmetrization effects are typically unimportant for $q\gtrsim 50$ MeV/c. Aside from resonant interactions, e.g. $\rho^{0}\rightarrow \pi^+\pi^-$, the strong interaction is most manifest at small relative momentum due to the lack of competing phase space. Other resonant interactions certainly provide peaks, but those are usually considered within the context of charge balance correlations. Coulomb interactions are relatively weak in magnitude, but extend over larger relative momentum. For large $q$ the squared wave functions can be considered classically \cite{Kim:1992zzc}, and when averaged over direction depend on $q$ as
\begin{eqnarray}
|\phi(\vec{q},\vec{r})|^2&\approx& 1.0-\frac{2\mu z_1z_2e^2}{q^2r}.
\end{eqnarray}
A classical expression also exists to account for the dependence on the angle between $\vec{q}$ and $\vec{r}$ \cite{Kim:1992zzc}, and is presented in Appendix \ref{app:classical}. The same- and opposite-sign correlation functions have oppositely signed contributions from Coulomb correlations. Thus, they reinforce one another when constructing a BF. For more central collisions, the correlation functions weaken due to the $1/r$ dependence above. However, when translating a correlation function to a BF, a factor of the multiplicity arises. The radii roughly scale as $(dN/dy)^{1/3}$, so the Coulomb contribution to the BF should increase with multiplicity, roughly as $(dN/dy)^{2/3}$. Thus, Coulomb effects might provide non-negligible contributions to the BF, even if their contribution to the correlation function is below a tenth of a percent. 

If two charged particles, with momenta $\vec{p}_a$ and $\vec{p}_b$ and charges $Z_a$ and $Z_b$, interact via the Coulomb interaction at large relative momenta, one can ask whether the interaction should be screened by the fact that both $a$ and $b$ are accompanied by balancing charges. For large relative momenta, which corresponds to large relative position, those balancing charges should perfectly screen the Coulomb interaction, because particle $b$ should see both $a$ and the balancing charge of $a$. In \cite{Pratt:2003gh} the screening effect was crudely estimated with a pion gas, and it was seen that without screening the Coulomb interaction noticeably distorted the BF, but that after accounting for screening the Coulomb effect only affected the first few bins of relative rapidity. Here, we improve on that picture by accounting for the fact that the balancing charges are spread across all species of particles. For example, the existence of a positive kaon not only promotes the existence of a negative kaon, but also promotes more or few pions, protons, or their antiparticles. Decays, which were neglected in the previous study, are taken into account here. Finally, in this study distortions from FSI are also calculated for $p\bar{p}$ and $K^+K^-$ BFs.

For the standard algorithm described above, an uncorrelated pair, $a,b$, is generated then weighted with $|\phi_{ab}(\vec{q},\vec{r})|^2$. The two-particle distribution is then assigned a weight, which, if the particle did not interact, would be unity. To include screening, one must alter the weight to include the interaction with accompanying particles,
\begin{eqnarray}
\label{eq:interpair}
C_{ab}-1&\approx& \left[|\phi_{ab}(\vec{q}_{ab},\vec{r}_{ab})|^2-1\right]
+\sum_{a'}K_{a'|a}\left[|\phi_{a'b}(\vec{q}_{a'b},\vec{r}_{a'b})|^2-1\right]\\
\nonumber
&+&\sum_{b'}K_{b'|b}\left[|\phi_{ab'}(\vec{q}_{ab'},\vec{r}_{ab'})|^2-1\right]
+\sum_{a'b'}K_{a'|a}K_{b'|b}\left[|\phi(\vec{q}_{a'b'},\vec{r}_{a'b'})|^2-1\right].
\end{eqnarray}
This expression accounts for all the interactions between the particle $a$ and its accompanying balancing cohort and the particle $b$ and its cohort. Final-state interactions within a cohort are ignored, aside from those that were responsible for the kernel $K$. For a given particle $a$ there are many more particles in other cohorts than in the same cohort. The indices $a,a',b,b'$ reference all the information of a specific particle including its type, momentum and position. The usual Koonin equation would ignore the latter three terms in Eq. (\ref{eq:interpair}). 

One might have chosen a different form for the correlation weight $C_{ab}$ in  Eq. (\ref{eq:interpair}). An obvious choice might be to take the product of the four wave functions rather than the sum. In the limit that the wave functions are near unity the choices become identical. For Coulomb or strong interactions, the variation of $|\phi_{ab}|^2$ from unity are indeed small except in a small region of phase space of $q_{ab}$, and the chance that for some sets of particles that multiple values of $q$ are not small, the two choices should be similar. For identical particle interference, the form of $|\phi|^2$ could be $1\pm\cos(qr)$. The oscillating piece is not small, but for most pairs $qr$ is large and the oscillations simply provide noise. Thus, the final answer should not be significantly dependent on exactly how Eq. (\ref{eq:interpair}) is chosen. 

When $q_{ab}$ is large, weights are dominated by Coulomb interactions. In this case the kernel weights combined with the fact that the factors $[|\phi|^2-1]$, which are proportional to the ratios of charges, should lead to a cancellation. Physically, this can be considered as screening. If the particle $b$ has a large relative momentum to $a$, one expects that the balancing cohort to $a$ should cancel the interaction. In contrast, for small relative momentum $a$ and $b$ would spend significant time under on another's influence, and the effects of the cohorts should disappear.

To generate the correlations described by Eq. (\ref{eq:interpair}) one needs to sum over all hadron species $a'$ and $b'$ that accompany $a$ and $b$. The particles $a'$ are first generated according to their final yields, i.e. they are chosen with probability $p'=\langle\langle N_{h'}\rangle\rangle/\langle\langle N\rangle\rangle$. The positions of $a$ and $a'$ are chosen in a correlated manner in the same manner described for calculating BFs in Sec. \ref{sec:algorithm_bf}. The additional weights, $W_{h'|h}$, defined in Eq. (\ref{eq:KW}) are used to modify the correlation weights in Eq. (\ref{eq:interpair}),
\begin{eqnarray}
C'_{ab}-1&\approx& \left[|\phi_{ab}(\vec{q}_{ab},\vec{r}_{ab})|^2-1\right]
+\sum_{a'}K_{a'|a}\left[|\phi_{a'b}(\vec{q}_{a'b},\vec{r}_{a'b})|^2-1\right]W_{a'a}\\
\nonumber
&+&\sum_{b'}K_{b'|b}\left[|\phi_{ab'}(\vec{q}_{ab'},\vec{r}_{ab'})|^2-1\right]W_{b'b}
+\sum_{a'b'}K_{a'|a}K_{b'|b}\left[|\phi(\vec{q}_{a'b'},\vec{r}_{a'b'})|^2-1\right]W_{a'a}W_{b'b}.
\end{eqnarray}
An array is calculated to represent the numerator of the correlation function. Based on the momenta of $a$ and $b$ the appropriate bin is chosen, then incremented by $C'_{ab}$. A separate array is used for the numerator, but it is incremented by unity. Finally, the correlation function is found by dividing the numerator's array by that of the denominator. The correlation for a given bin thus represents the average of $C'_{ab}$ for pairs, $a,b$, that fit that bin. 

In some cases the particles $a$ and $b$ described above are unstable, i.e. they decay after being emitted from the fireball which is chosen for any decays with lifetimes greater than 100 fm/$c$.  This might include long-lived states like the $\eta$ meson. In that case the weight $C'_{ab}$ described above is used to increment the bins defined by any decay products of $a$ with and decay products of $b$. 


\section{Results: Distortions to BFs from Final-State Interactions}\label{sec:results_femtodistortions}

Femtoscopic correlation functions were first generated with the blast-wave model. Blast-wave parameters were chosen to fit the spectra and pion source sizes. The fitting procedure for choosing the blast-wave parameters is described in Appendix \ref{app:blastwave}. They were $T_c=150$ MeV, $T_f=100$ MeV, $U_\perp=1.092$, $R=13.4$ fm and $\tau=13.4$ fm/$c$. It should be emphasized that the blast-wave model is crude. Fitting to blast-wave models tends to result in shorter breakup times than seen in much more realistic hybrid models which incorporate both a hydrodynamic stage and a microscopic hadronic simulation. However, these parameters do roughly reproduce both the spectra and like-sign pion femtoscopic correlations, so they are well suited for the purpose of this study, which is to gauge the importance of these effects in BF analyses. Additionally, parameters were chosen to represent the spread of the charge correlation in coordinate space, $\sigma_\eta=0.5$ and $\sigma_R=3.0$ fm. These last two parameters crudely reproduce experimental BFs, but with the emphasis on being crude. The spread should be significantly broader for $p\bar{p}$ and $K^+K^-$ BFs than for $\pi^+\pi^-$ BFs. Nonetheless, for gauging the effect of femtoscopic correlations on BFs, a rough picture is sufficient. The calculations presented here required a large amount of statistics due to the small size of the effect and the noise related to the inclusion of cross-correlations from balancing charges. The number of pairs generated for the calculations here exceeded $10^{12}$, which would have made using a more sophisticated, and slower, model untenable.

\begin{figure}
\includegraphics[width=0.48\textwidth]{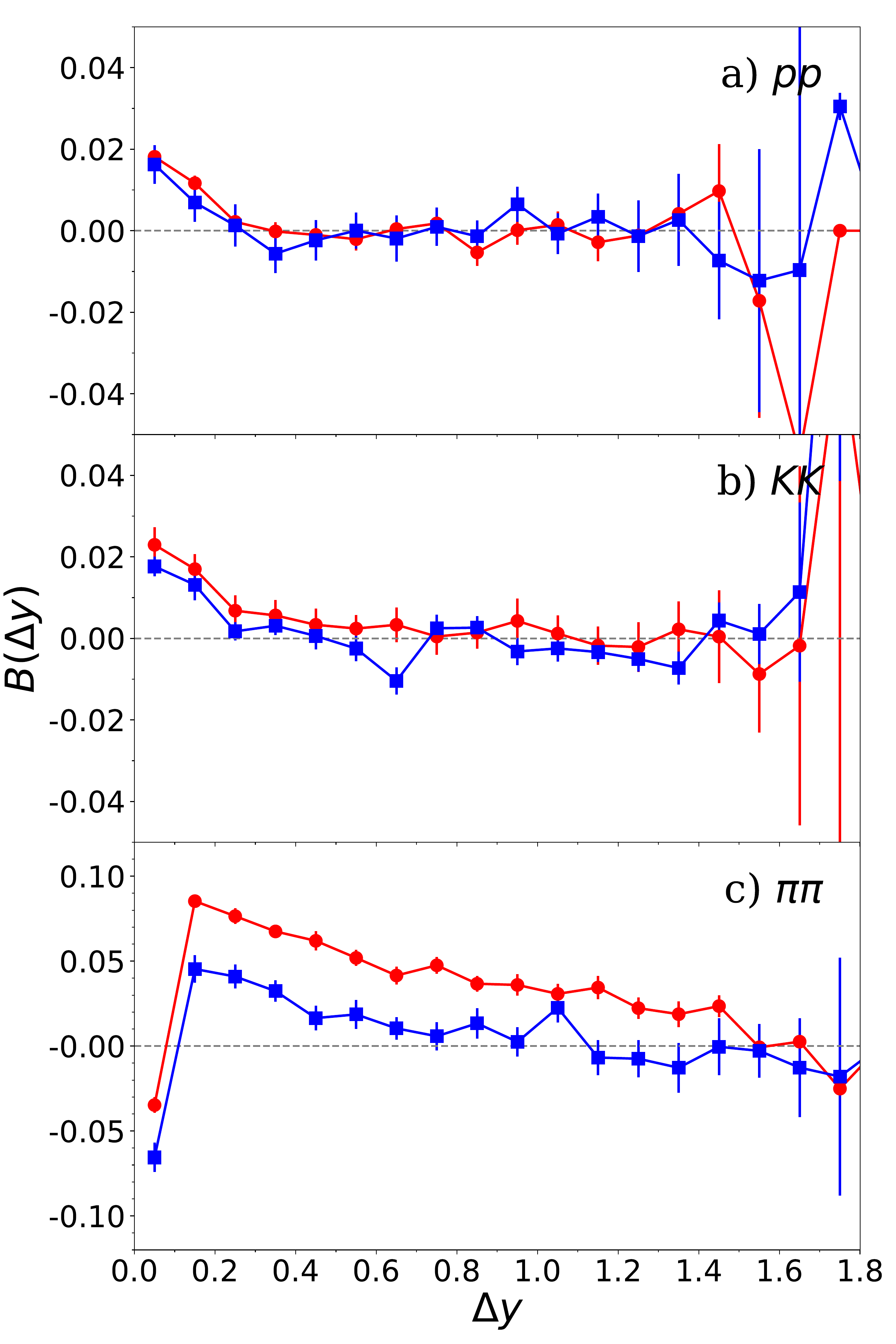}
\includegraphics[width=0.48\textwidth]{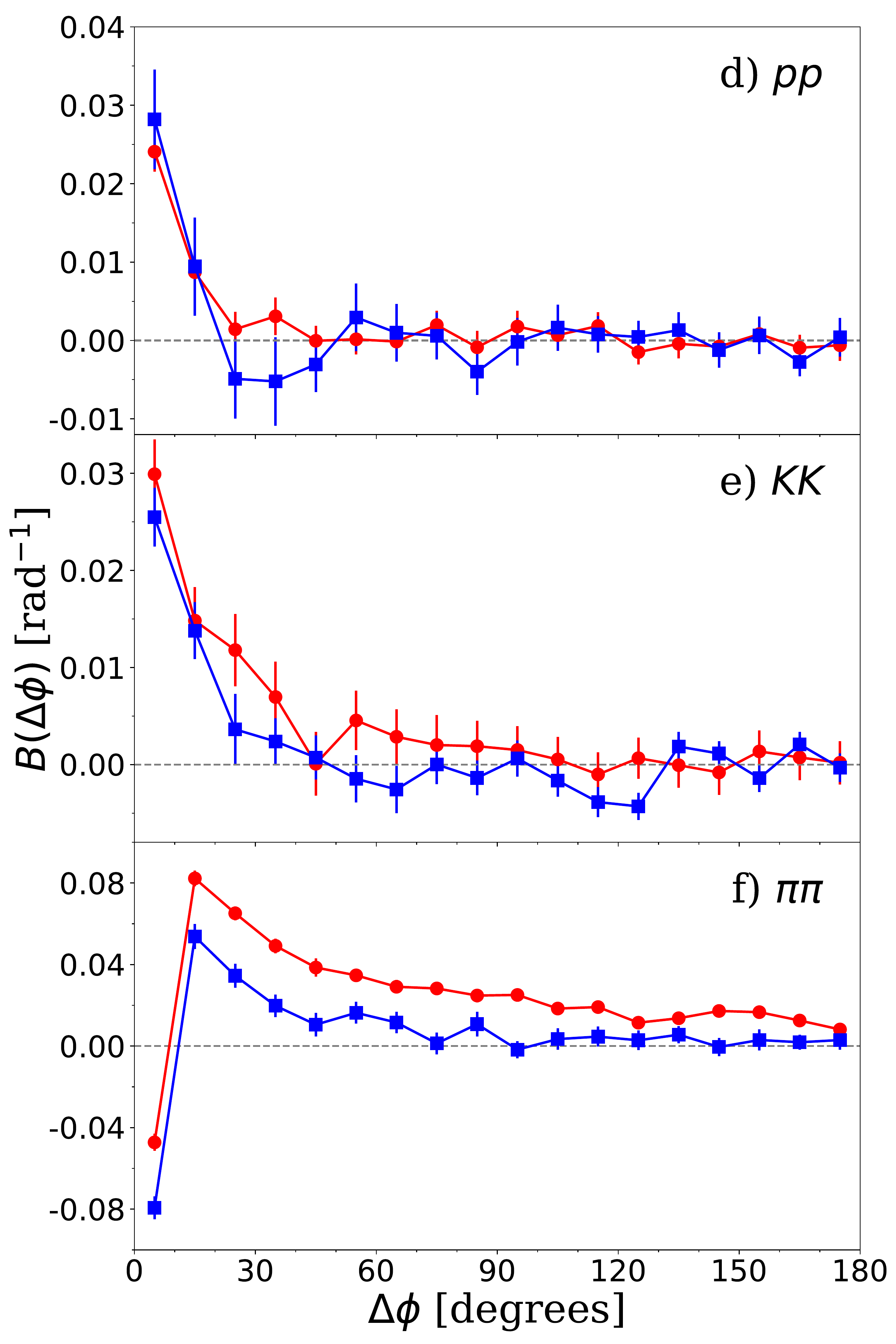}
\caption{\label{fig:bfhbt}
Contributions to BFs from femtoscopic correlations are shown as a function of relative rapidity in panels {\it a} ($pp$), {\it b} ($KK$) and {\it c} ($\pi\pi$). The kaon and proton BFs are affected marginally, and only in the first two bins. The contribution never exceeds more than 0.02. The contribution to the $\pi\pi$ BFs are more substantial and extend further in rapidity. The femtoscopic contributions are displayed with (red circles) and without (blue squares) screening. The screening mainly affects results at larger relative rapidity, and significantly lowers the femtoscopic contribution to the $\pi\pi$ BF. The right-side panels, {\it d-f}, show the same behavior when binned by relative azimuthal angle. 
}
\end{figure}

Using the methods described in Sec. \ref{sec:theory_screening}, femtoscopic correlations were calculated. To reduce noise in the femtoscopic correlations below a tenth of percent, over a trillion pairs were analyzed. Because particles were sampled according to their multiplicities, correlations for kaons or protons were noisier than for pions. Figure \ref{fig:bfhbt} shows the contribution to BFs from FSI. Calculations are displayed both with and without screening. For results without screening correlation functions were calculated using Koonin's equation, Eq. (\ref{eq:koonin}), which neglects how FSI between two particles affect correlations those other particles involved in balancing the charges of the first two.  If not for screening, a non-negligible contribution would be present in the $\pi-\pi$ BFs and extend to larger relative rapidity or relative azimuthal angle. BFs were generated by multiplying regular correlation functions by the multiplicity of uncorrelated particles in the same bins. Because pions have higher multiplicity, the effect on the BFs was more noticeable. After the inclusion of screening the distortion to the BFs are only in the first few bins, at small relative rapidity or angle. For $\pi\pi$ the contributions in the first bin are negative due to the positive contribution from the same-sign correlation function due to identical-particle statistics. For slightly larger relative momentum femtoscopic effects are mainly from the Coulomb interaction. The Coulomb contribution to the correlation functions are negative for same-sign correlations and positive for opposite-sign correlations. The BF contribution, which is constructed by subtracting the same-sign correlation from the opposite-sign correlation, is positive.

\begin{figure}
\includegraphics[width=0.48\textwidth]{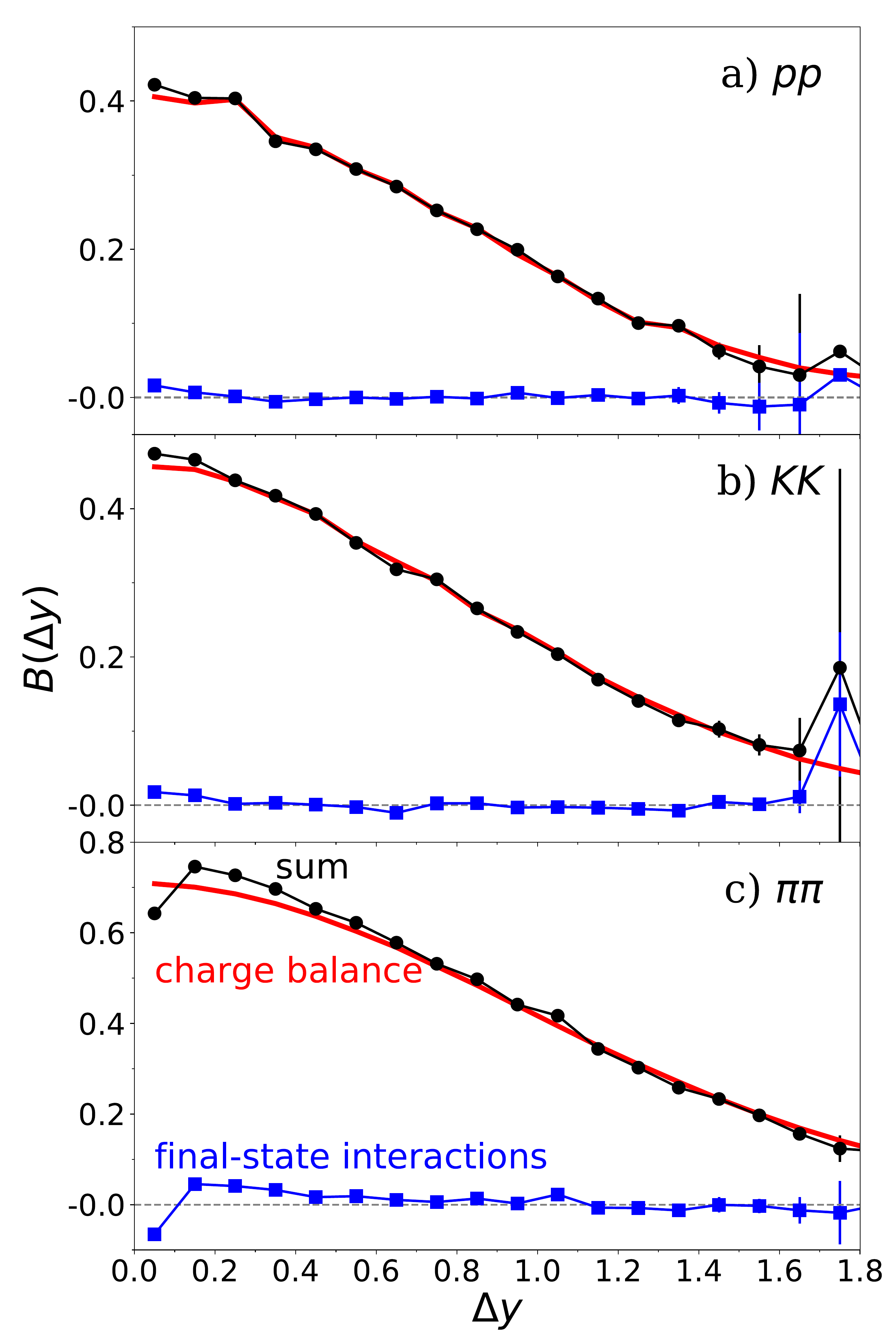}
\includegraphics[width=0.48\textwidth]{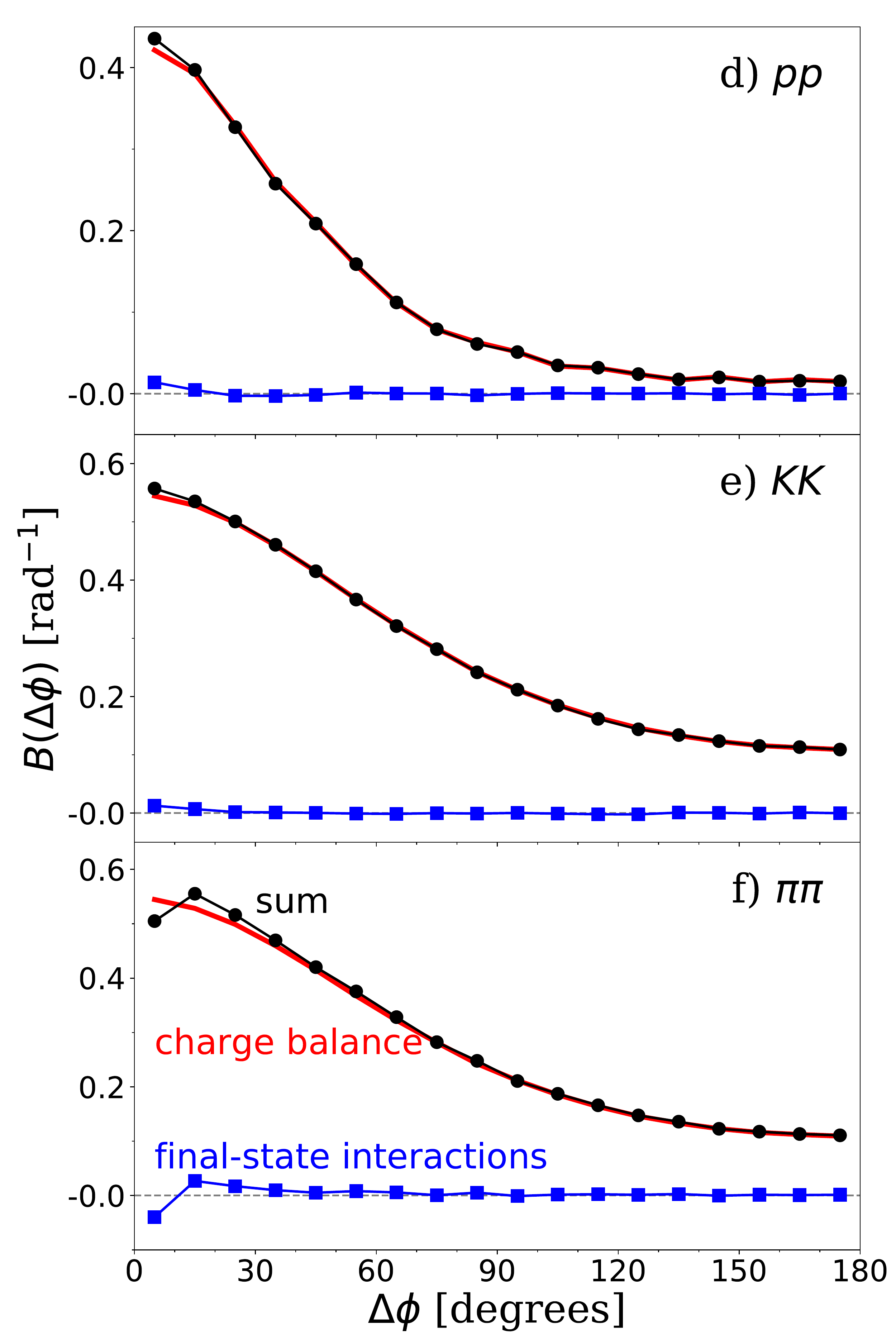}
\caption{\label{fig:bftotal}
The contribution to the BF (blue squares) is shown against the simple BF from local charge conservation only (red line), along with the sum of the two contributions (black circles). The effects of femtoscopic correlations are modest, but noticeable, for the $\pi\pi$ BFs. The dip at small relative momentum derives from the positive correlation from identical-particle interference in the same-sign correlation functions. The slight positive corrections is due to final-state Coulomb interactions. The same behavior is seen in BFs binned by relative rapidity (left-side panels) and relative azimuthal angle (right-side panels).
}
\end{figure}

To gain insight into whether the distortions to the BF from FSI are significant, the femtoscopic contribution, with screening included, is added to the main contribution from charge balance. The calculation for the main charge balance is a rather crude model, and should not be taken seriously to better than 10-20\%, but is sufficient for gauging the relative strength of the femtoscopic contributions. Calculations of the BFs with and without the femtoscopic contributions are displayed in Fig. \ref{fig:bftotal}.

Figures \ref{fig:bfhbt} and \ref{fig:bftotal} address the first questions posed for this study. BFs are modified slightly, but noticeably, by femtoscopic correlations. Those contributions are mainly in the first several bins of relative rapidity or relative azimuthal angle. Whereas the $\pi\pi$ BFs are noticeably affected, albeit modestly, the modifications to the $pp$ and $KK$ BFs are negligible. The Coulomb contribution to the femtoscopic correlation functions are of similar magnitude for $\pi\pi$, $KK$ and $pp$ correlations, but BFs involve multiplying correlations by the multiplicity of background pairs in a given bin, which is a significantly smaller factor for protons and kaons. Thus, it was not surprising that the effects are larger for $\pi\pi$ balance functions. 

The shape of the modification for the $\pi\pi$ BF was also as expected as it was seen in \cite{Pratt:2003gh}. The magnitude of the effect is reduced compared to the calculation in \cite{Pratt:2003gh}, but that calculation had ignored the effect of long-lived decays, which reduces the magnitude of the femtoscopic correlation. The dip for the bins with lowest relative rapidity or angle was due to identical-particle interference for same-sign pions. The rise for the next few bins is due to the Coulomb interaction. As shown in Fig. \ref{fig:bfhbt} this part of the effect was significantly dampened by the inclusion of screening effects. The fact that each charge is accompanied by balancing charge of the opposite sign effectively screens the charge, unless the relative momentum is so small that the screening charges have little chance of standing between the charges of interest. If the calculations had been performed at lower beam energy, Coulomb effects would have been smaller. This is because Coulomb forces are long range and thus a given charge affects a greater number of other charges when there are more charges present.

By accounting for the FSI weights of balancing particles, the distortions to the BFs from Coulomb effects is significantly muted. Further, by applying these weights to and from balancing partners, the correct normalization was restored. Even for FSI from identical particles, the normalization would have been incorrect if only the Koonin contribution to the BFs had been considered. For identical-particle statistics, symmetrization affects only those other pions within a similar bin of phase space, a number which is set by the local phase space density. Thus, if the average phase space density if 5\%, there tends to be an overall enhancement of 0.05 to the area underneath the BF. If the calculations were repeated for less central collisions, the net contribution to the BF from symmetrization would be similar, but it would be spread out over larger relative momentum because larger source sizes lead to more extended correlation functions. The dip for small $\Delta y$ and small $\Delta\phi$ would then be less pronounced.

One clear result of these calculations is that FSI distortions are nearly negligible for $pp$ and $KK$ BFs. This is important because those BFs play crucial roles in understanding the chemical evolution and diffusivity of the matter created in heavy-ion collisions.


\section{Results: Distortions to Femtoscopic Correlations from Charge-Balance Correlations}\label{sec:results_bfdistortions}

The effect of charge-balance correlations are typically neglected in calculations of correlations functions for femtoscopic purposes. Here, we investigate the degree to which that is justified. First, femtoscopic correlations were calculated from the blast wave model as described in Appendix \ref{app:blastwave}. Correlations were found for both $\pi^+$ and $\pi^-$ pairs. BFs were then calculated for the simple parametric model described in Appendix \ref{app:blastwave}. The difference between the like-sign and opposite-sign correlations from BFs is then
\begin{eqnarray}
C_{\rm opp.~sign}(Q_{\rm inv})-C_{\rm same~sign}(Q_{\rm inv})&=&
\frac{1}{dN_{\pi}/dQ_{\rm inv}}B(Q_{\rm inv}),\\
\nonumber
Q_{\rm inv}^2&\equiv& -(p_1-p_2)^2.
\end{eqnarray}
Here, $dN_{\pi}/dQ_{\rm inv}$ is the number of pion pairs of the same sign separated by $Q_{\rm inv}$ divided by the number of pions of that same sign. 

\begin{figure}
\centerline{\includegraphics[width=0.6\textwidth]{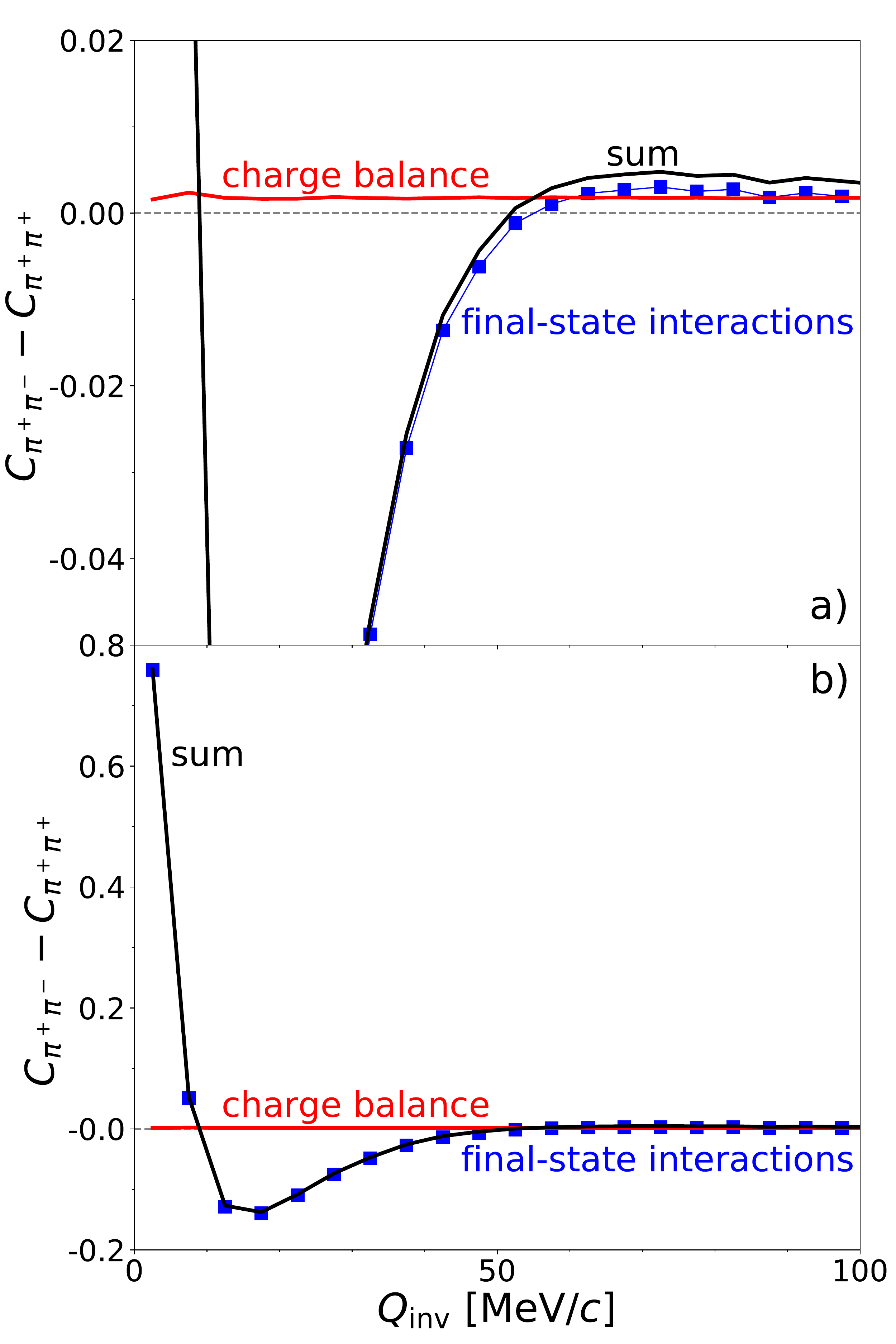}}
\caption{\label{fig:cfqinv}
The contribution of charge-balance effects (red line) is much smaller than the femtoscopic correlation (blue squares) at small relative momentum. The net correlation is thus little changed. Because BFs provide the difference between opposite-sign and same-sign correlations, and because the effect was small, only the difference between $C_{\rm opp.~sign}$ and $C_{\rm same~sign}$ correlations were analyzed. The same calculations are shown in both panels, with the vertical scale in panel (a) being magnified to show the size of the contribution from final-state interactions.
}
\end{figure}

Figure \ref{fig:cfqinv} displays femtoscopic correlations alongside those for BFs. The factor $dN_\pi/dQ_{\rm inv}$ scales as $Q_{\rm inv}^2$ at low $Q_{\rm inv}$ due to phase space. For this reason the effect of charge balance is muted at low relative momentum, and the effect never rises above the half-percent level. This level of distortion is negligible given the current precision with which identical-pion femtoscopy is being analyzed. Because BFs are constructed by taking the difference between opposite-sign and like-sign correlations, it is difficult to assign that correlation specifically to either $C_{\rm same~sign}$ vs $C_{\rm opp.~sign}$. For charge balance from decays late in the reaction, one expects most of that strength to appear in the opposite-sign correlation. However, charge balance correlations from equilibrated systems, before final decays, tends to be split evenly between the opposite-sign and same-sign pieces if the systems are large \cite{Pratt:2020ekp,Savchuk:2019xfg}. Luckily, given that the contributions are so small, it does not matter what fraction of it should be assigned to the same-sign vs. opposite-sign correlation functions. 

The main lesson taken from Fig. \ref{fig:cfqinv} is that femtoscopic analyses can safely ignore the contributions from charge balance for central heavy-ion collisions. For peripheral collisions or for $pp$ collisions, the effects are probably non-negligible. For small source sizes femtoscopic correlations can extend to $Q_{\rm inv}\sim 200$ MeV/$c$ and $dN_{\rm \pi}/dQ_{\rm inv}$ can be small. Also, for small systems other classes of correlations also tend to interfere with the result, including momentum conservation. In fact, the validity of the Koonin equation comes into question when the overall source size is not much larger than the inverse characteristic momentum \cite{Pratt:1997pw}.


\section{Summary}\label{sec:summary}

BFs represent the best means for addressing questions about chemical evolution in high-energy heavy-ion collisions. In particular, one needs to evaluate the shape of the BF when binned by rapidity. If the quark chemistry is equilibrated within the first fm/$c$, balancing charges can separate by $\sim 1$ unit of spatial rapidity by the time hadrons are finally emitted from the fireball. This is manifested by broad BFs, particularly for $K^+K^-$ and $p\bar{p}$ BFs. However, two other classes of phenomena also provide correlations that might potentially interfere with the interpretation of BFs. The first is correlation from final-state interactions, which represents the topic of this paper. The second is baryon-baryon annihilation, which is a topic for future study.

The contribution of femtoscopic correlations, i.e. those from final-state interactions, was estimated in a previous study. But for that study, only pions were considered, long-lived decays were neglected, and the distortions of BFs binned by relative azimuthal angle were not considered. Given the importance of the shapes of the $K^+K^-$ and $p\bar{p}$ BFs, it was felt that a new study was needed. In the basic formulation, i.e. the Koonin formula, femtoscopic correlations enhance the emission of like-sign pions due to the symmetrization of the two-particle outgoing wave function. This provides a negative contribution to the BF. Coulomb effects enhance the emission of opposite-sign pairs, whereas they discourage the emission of same-sign pairs. For $pp$ or $\bar{p}\bar{p}$ pairs, a resonant-like interaction at small relative momentum enhances the emission of same-sign pairs. However, the net integral of the BF must be unchanged, because for every extra particle of a given charge, there must exist exactly one extra particle of the opposite sign, regardless of FSI. If the emission of same-signed pairs is enhanced by some effect then the emission of opposite-sign pairs must also be correspondingly enhanced to maintain the strict requirement of global charge conservation. 

The issues described in the previous paragraph motivated the current study. An ambitious model was developed where additional weight from final-state interactions was applied not only to the interacting pair, but to any balancing partners. This required modeling how each charge particle was accompanied by additional particles. For each charged particle $a$ of hadron type $h$, a probability was found for it to be accompanied by a hadron of type $h'$. The additional hadron $a'$ was then placed in vicinity of $a$ according to a parametric form of the correlation. The charge-balance arguments from Sec. \ref{sec:theory_bf} show how one can consider $a'$ as being any hadron, then applying a balancing weight $w(a'|a)$ based on charge balance. The weight $w(a'|a)$ takes into account charge balance at the point of chemical equilibrium and decays to determine how the weight depends on the the specific species $a$ and $a'$. The correlation of $a$ and $a'$ in momentum space was crudely modeled by assuming a simple correlation in coordinate space that is mapped onto momentum space via a blast-wave model. In addition to parameters to set the temperature and flow velocity, the blast wave model had parameters describing how the emission points of $a$ and $a'$ would be correlated in coordinate space. If one is considering the interaction weight of pion $a$ with pion $b$, one must also apply that weight to all the balancing partners of $a$, i.e. those denoted by $a'$, with $b$ and all its balancing partners of $b'$. Because the charge of the balancing partners $a'$ exactly cancel those of $a$, the interaction weight for $a$ and $b$ is also applied to opposite sign pairs, albeit spread over a wider range of relative momentum. This preserves the charge conservation constraint of the BF in a way that more realistically accounts for how balancing charge is spread amongst different species at different locations. Additionally, weights were projected through the chain of decays occurring between a point where chemical occurred and when the particles are emitted. This rather long-winded procedure is especially necessary for Coulomb interactions. Once a particle $b$ is separated from $a$ by larger relative momentum, it is as likely to feel the interactions with the balancing particle $a'$ as it is to be be influenced by $a$. Thus, the balancing charge effectively screens the Coulomb effects for larger relative momentum. 

The approach and methods described and developed herein were then applied to calculating the femtoscopic contributions to $\pi^+\pi^-$, $K^+K^-$ and $p\bar{p}$ BFs. Significant effects were only found for the $\pi^+\pi^-$ case. Although the effect on correlation functions is of similar strength for all three cases, the translation to BFs involves a factor of the multiplicity, which is higher for pions than for kaons or protons. The contribution to the $\pi^+\pi^-$ BF was confined to the first few bins in relative rapidity or azimuthal angle, but would have extended further if  screening effects had not been included. The size of the correction for conditions similar to central collisions of Au+Au at RHIC were modest and somewhat smaller than what was found with the simpler model considered in \cite{Pratt:2003gh}.

The main conclusions of the study are that
\begin{enumerate}\itemsep=0pt\vspace*{-8pt}
\item Femtoscopic correlations should provide a modest dip in the $\pi^+\pi^-$ BF at small relative rapidity or relative azimuthal angle, followed by a small enhancement at slightly larger values.
\item For $K^+K^-$ or $p\bar{p}$ BFs, the effect of correlations from final-state interactions is negligible.
\item Correlation functions at small relative momentum used for femtoscopic purposes based on final-state interactions can safely neglect the influence of charge-conservation effects, at least for central heavy-ion collisions.
\end{enumerate}

These findings are reassuring. They validate the practice of treating femtoscopic and charge-balance effects separately, although one might wish to apply a small FSI correction to $\pi^+\pi^-$ BFs. The rather crude nature of the modeling here, especially the use of a blast-wave, should predict this additional structure to the $\sim 10\%$ level, but give that the distorting effects are at the five percent level, calculating the distortion of a 5\% effect to ten percent accuracy should be sufficient to add the corrections from a simple model to BF calculations from more sophisticated models.

As mentioned earlier, there is an additional effect that might also complicate the interpretation of BFs. Baryon annihilation depletes the $pp$ BF at smaller relative momentum, relative rapidity or relative angle. Combined with this study, a detailed estimate of how annihilation affects BFs should enable the confident interpretation of experimental BFs. This is crucial if BFs are to provide a quantitative and rigorous means for extracting information about the chemistry and diffusivity of matter created in relativistic heavy-ion collisions.

\appendix
\section{Blast Wave Model}\label{app:blastwave}

Charge conservation correlates balancing particles in coordinate space. The correlation is then projected onto momentum space through collective flow. A blast wave model provides a simple parametric means to describe final-state collective flow. For this study, a particularly simple blast wave prescription is applied. Particles are all emitted at a fixed proper time $\tau_f$. This is the time measured by an observer moving with a constant velocity from the $z=0$ plane at time $t=0$ to the emission point. In terms of the laboratory time $t$ and the longitudinal coordinate $z$,
\begin{eqnarray}
\tau&=&\sqrt{t^2-z^2}.
\end{eqnarray}
In terms of spatial rapidity,
\begin{eqnarray}
\eta_s&=&\frac{1}{2}\ln\left(\frac{t+z}{t-z}\right),
\end{eqnarray}
emission is given a Gaussian distribution corresponding to the finite rapidity range of emission at RHIC,
\begin{eqnarray}
dN/d\eta_s&\sim& e^{-\eta_s^2/2\Sigma_\eta^2},
\end{eqnarray}
with $\Sigma_\eta=1.8$.

The distribution of emission points in the transverse plane is considered a constant up to some maximum radius, $R$. The momenta is determined by a temperature $T_f$ and a transverse collective velocity parameterized by $u_\perp$,
\begin{eqnarray}
u_i&=&u_\perp\frac{r}{R}.
\end{eqnarray}
Particles were generated stochastically. Final yields were scaled to reproduce the experimental number, so the blast wave model only serves a a means to assign momenta and space-time coordinates to the momenta. Species were chosen  proportional to the multiplicity at the time of emission, $\langle\langle N_h\rangle\rangle$. This multiplicity was determined by first generating particles proportional to their densities in an equilibrated system at temperature $T_c=150$, corresponding to the densities latest time at which chemical equilibrium might have been maintained, $\langle n_h\rangle$. Particles were then decayed according to their branching ratios. All decays with lifetimes less than 100 fm/$c$ were simulated. The products were then randomly placed in the blast-wave volume and assigned momenta consistent with the final blast wave temperature and collective velocity. Any further decay was simulated. 
\begin{eqnarray}
\langle\langle N_h\rangle\rangle&=&
\sum_{H,c_H}
\langle N_H\rangle b_{c_H}m_{c_H,h}.
\end{eqnarray}
where $b_{c_H}$ is the branching ratio for a particular channel $c_H$ and $m_{c_H,h}$ is the number of hadrons of type $h$ in that channel. This prescription does ignore the fact that some short lived particles, like $\Delta$ baryons or $\rho$ mesons, might still exist at the final breakup. Though the number of such resonances should be significantly fewer as compared to the earlier equilibrium, regeneration would suggest that a number of such resonances would be emitted at the final time with all the decay products escaping rescattering. But this should have little effect on spectra because most of the resonances are rather broad so that the final momenta differ only slightly compared to being re-thermalized. Further, because the lifetimes are short, femtoscopic correlations are not strongly affected.

\begin{figure}[H]
  \centering
  \includegraphics[scale=.5]{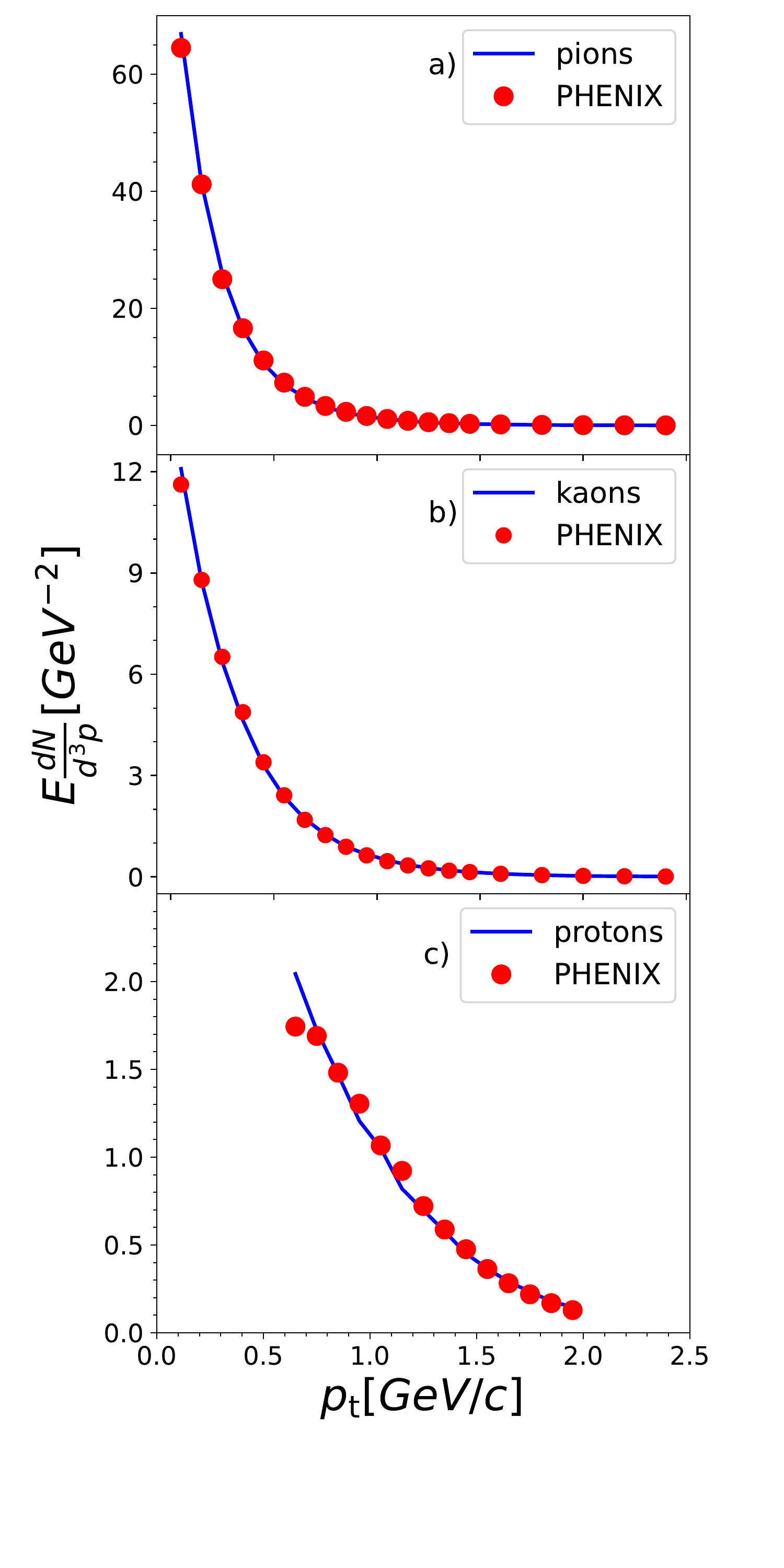}
  \caption{\label{fig:spectra}
  Spectra for pions, kaons and protons are compared to experimental results for central (0-5\% centrality) collisions of $200A$ GeV Au+Au collisions as a function of transverse momentum. Model results (blue lines) roughly match PHENIX results (red circles).}
\end{figure}

\begin{figure}[H]
  \centering
  \includegraphics[scale=.5]{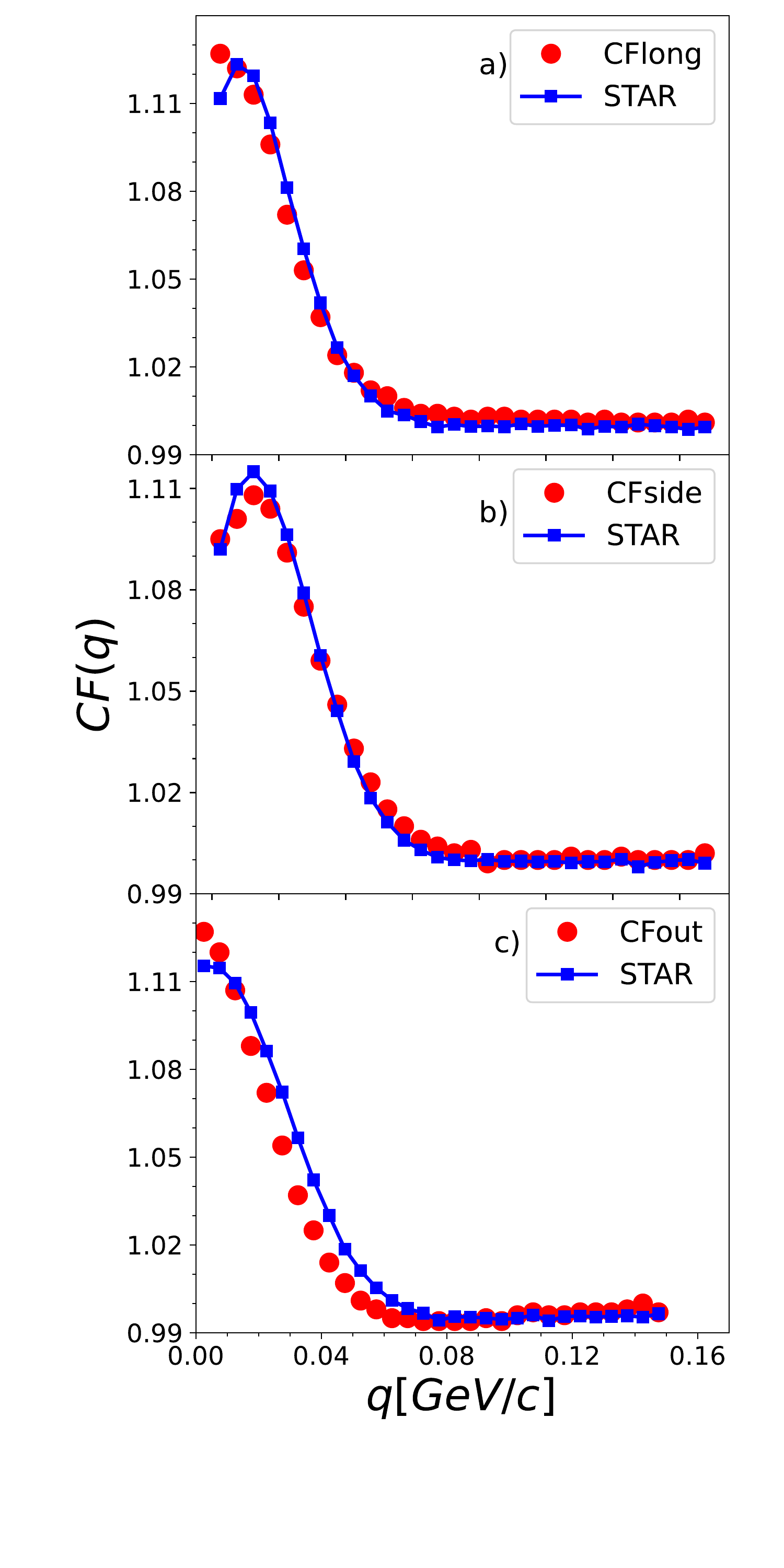}
  \caption{\label{fig:outsidelong}
  Two-pion correlation function projections as a function of relative momentum are shown for the model (blue lines) fit to data from the STAR Collaboration (red circles). Measurements are from $200A$GeV Au+Au collisions in the 0-5\% percent centrality range. The three projections are for relative momentum along the beam axis (``CFlong''), parallel to the pair momentum in the longitudinal comoving frame (``CFout''), and perpendicular to both the pair momentum and the beam axis (``CFside'').}
\end{figure}

Blast-wave parameters $T_f$, $R$, $\tau_f$, and $U_\perp$ were reproduced through comparison of simulated models with experimental data from $200A$ GeV Au+Au collisions at RHIC. For the spectra calculations MCMC generated hadrons were used to construct spectra, which were then compared to experimental data from the PHENIX Collaboration \cite{A.Adare:2013xfg}. A $\chi$-square minimization using the software describe in \cite{Virtanen:2019joe} was applied to obtain the most-likely parameters which are listed in Sec. \ref{sec:results_bfdistortions}. Fits are shown in Fig. \ref{fig:spectra} for kaons, protons and for pions. Modeling spectra produced a fit of the parameters $T_f$ and $U_\perp$ which were consequently utilized in the calculation of correlations to evaluate the final two parameters, the transverse size $R$ and the freeze-out time $\tau_f$. To generate the correlation functions, values of $\tau_f$ and $R$ were used to generate CFs using the Koonin prescription. CFs were then compared to experimental data from same-sign two-pion correlations functions measured by the STAR Collaboration \cite{J.Adams:2005xfg}. The same minimization used to fit the spectra was utilized to minimize the difference between data and experiment while varying the parameters of interest, with the best fit illustrated in Fig. \ref{fig:outsidelong}. The final parameter values are mentioned in Sec. \ref{sec:results_bfdistortions}.

\section{Classical Expressions for the Squared Coulomb Wave Function}\label{app:classical}

Here, we provide a slightly different form of the expressions derived in \cite{Kim:1992zzc}. The relation between the squared outgoing wave function and classical trajectories is
\begin{eqnarray}\label{eq:coulphasespace}
|\phi(q,r,\cos\theta)|^2_{\rm classical}&=&\left|\frac{d^3q_0}{d^3q}(q,r,\cos\theta)\right|\\
\nonumber
&=&\frac{q_0}{q}\left|\frac{d\cos\theta_0}{d\cos\theta}\right|.
\end{eqnarray}
Here, $\vec{q}$ is the asymptotic relative momentum whereas $\vec{q}_0$ is the relative momentum at the time of emission, when the separation was $\vec{r}$. The angle $\theta$ is between the vectors $\vec{q}$ and $\vec{r}$. Energy conservation, $q^2/2\mu=q_0^2/2\mu+Z_1Z_2e^2/r$, or equivalently $qdq=q_0dq_0$, was used to simplify the expression. Thus, $|\phi|^2_{\rm classical}$ describes how a phase space element $d^3q_0$ is focused into $d^3q$. 

To calculate the Jacobian, we consider a particle of mass $\mu$ at position $\vec{r}=r\hat{z}$ has an initial direction defined by $\theta_0$ and a final direction described by $\theta$. From \cite{Kim:1992zzc} one can see that conservation of angular momentum, energy and the Lenz vector allow one to express $\cos\theta_0$ in terms of $\cos\theta$,
\begin{eqnarray}\label{eq:epsdef}
\cos\theta_0&=&\frac{q}{q_0}\cos\theta-\frac{q}{2q_0}\frac{\epsilon}{(1\pm\sqrt{1-2\epsilon/(1+\cos\theta)})},\\
\nonumber
\frac{q_0}{q}&=&\sqrt{1-\epsilon},\\
\nonumber
\epsilon&=&\frac{Z_1Z_2e^2/r}{q^2/2\mu}.
\end{eqnarray}
Thus, $\cos\theta_0$ can be expressed solely in terms of $\cos\theta$ and $\epsilon$, the ratio of the initial Coulomb energy to the total energy in the center-of-mass frame. For when the charges have opposite sign, the interaction is attractive and $\epsilon<0$, whereas $\epsilon>0$ for same-sign pairs.

\begin{figure}
\centerline{\includegraphics[width=0.5\textwidth]{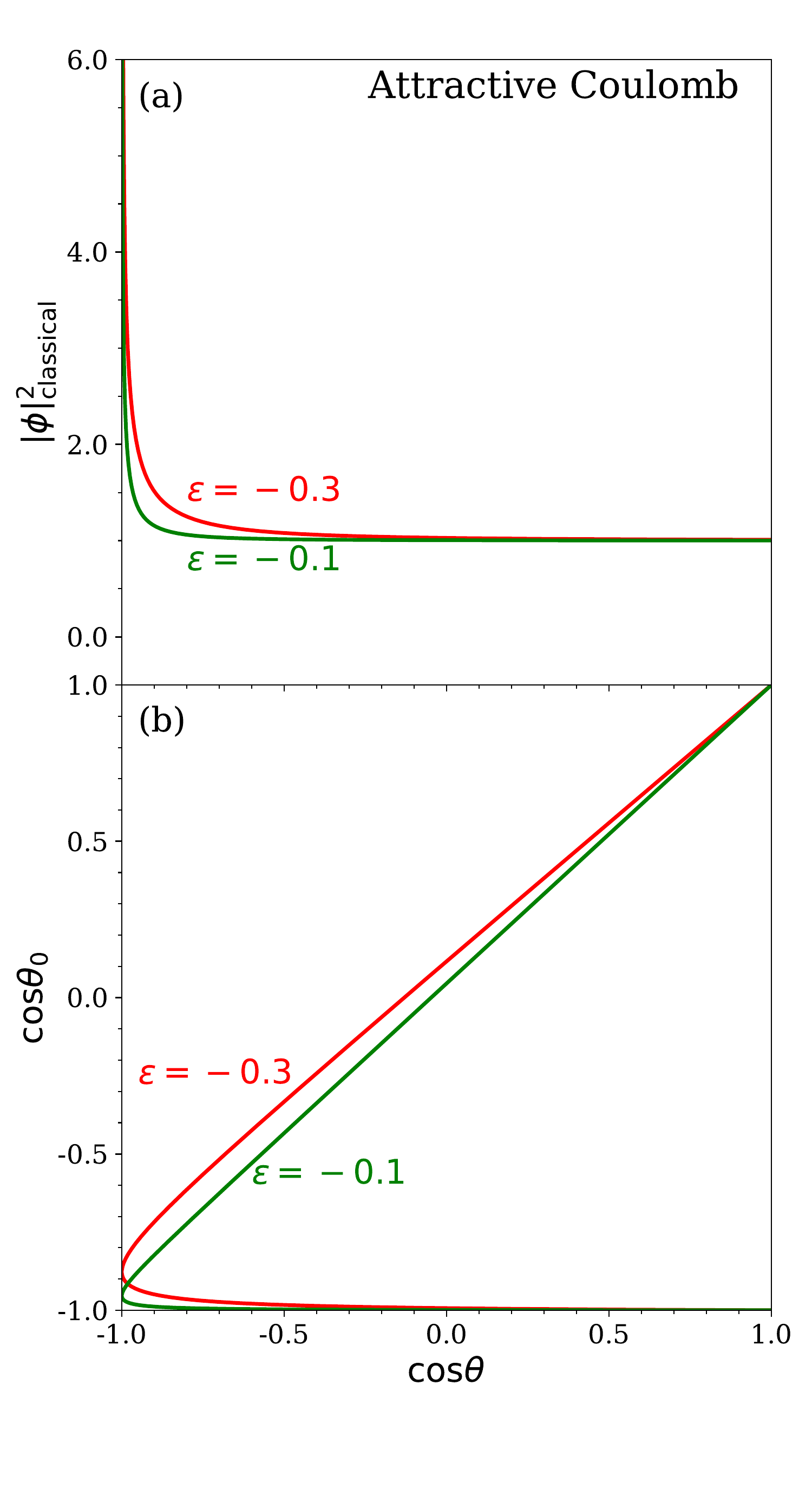}
\includegraphics[width=0.5\textwidth]{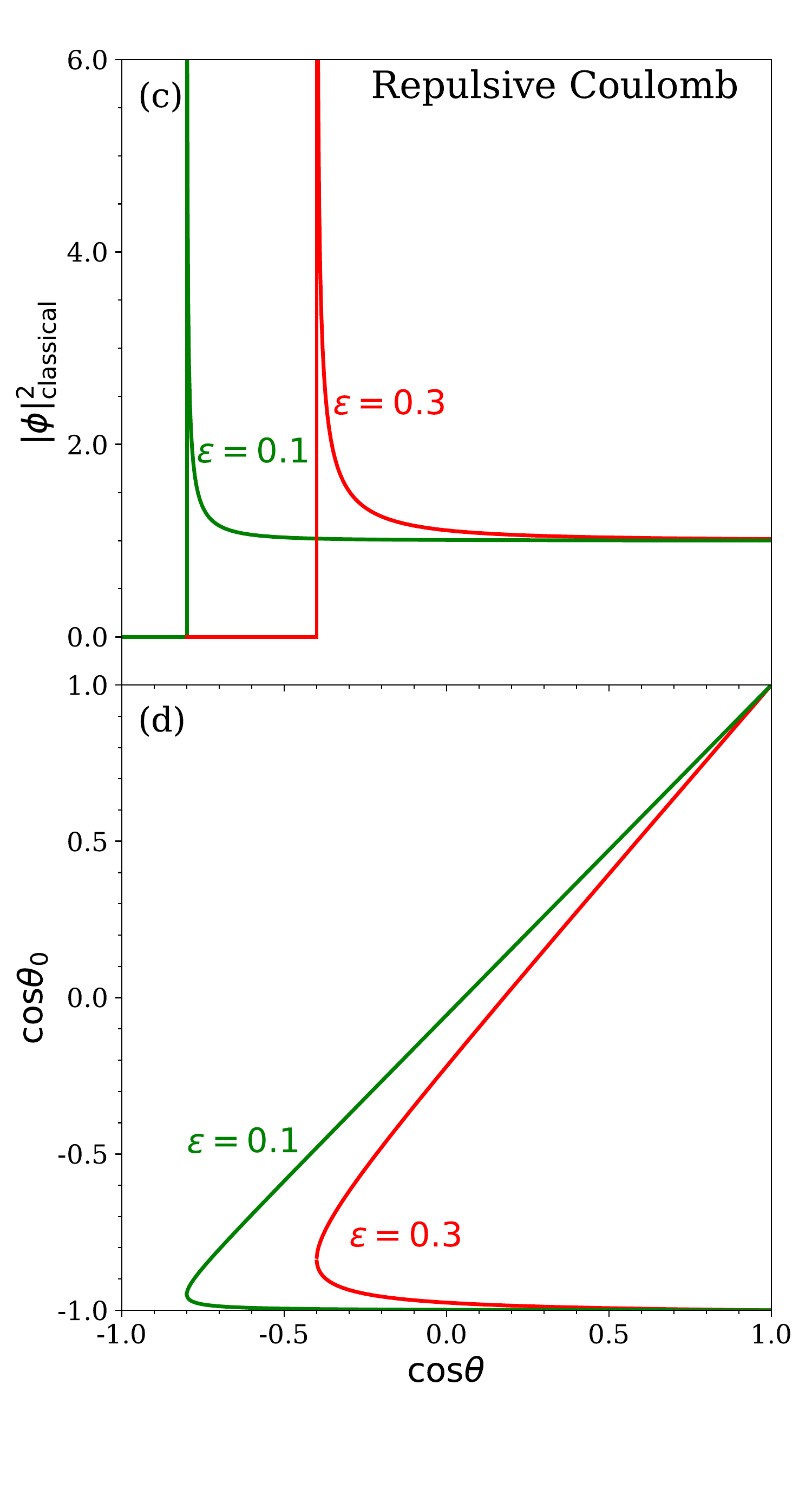}}
\caption{\label{fig:classical}
For large relative momenta, $Q_{\rm inv}>500$ MeV/$c$, classical expressions are applied for the squared relative wave functions. The case of attractive Coulomb interactions is displayed in panels (a) and (b), while the repulsive case is illustrated in (c) and (d). The classical analogy of the squared wave functions are generated from the relation between the final direction of the relative momenta and the initial direction, Eq. (\ref{eq:phi2class}. The mapping of the initial to the final direction of the momenta are shown in the lower panels of both figures where angles are relative to the original relative position. Because the squared wave function depends on the Jacobian, $d\cos\theta_0/d\cos\theta$, there are integrable poles in the wave function. The effective squared wave function, $|\phi|^2$, is completely determined by the final direction $\theta$ and $\epsilon$, defined in Eq. (\ref{eq:epsdef}).}
\end{figure}

Calculating $d\cos\theta_0/d\cos\theta$ and applying Eq. (\ref{eq:coulphasespace}) then gives the ``classical'' squared wave function,
\begin{eqnarray}\label{eq:phi2class}
|\phi(q,r,\cos\theta)|^2_{\rm classical}&=&\sum_{\pm} 1\pm\frac{1}{\beta}\left[\frac{\epsilon}{(1\pm \beta)(1+\cos\theta)}\right]^2,\\
\nonumber
\beta&=&\sqrt{1-\frac{2\epsilon}{1+\cos\theta}}.
\end{eqnarray}
There are two solutions to the trajectories, because there are two initial angles that can reproduce a given final angle. To understand the relation for $|\phi|^2_{\rm classical}$ it is useful to view the relationship between $\cos\theta$ and $\cos\theta_0$, which are illustrated for the attractive and repulsive cases in Fig \ref{fig:classical}. For the repulsive case, there are final angles which are unreachable, because the Coulomb force diverts  those trajectories with $\cos\theta_0$ near $-1.0$. In both cases, there are points for which $d\cos\theta_0/d\cos\theta$ are divergent, but these divergences are integrable. Even though there are divergences as $|\phi|^2_{\rm classical}\rightarrow\infty$ for the repulsive case, if one averages $|\phi|^2_{\rm classical}$ over $\cos\theta$, the result is below unity and
\begin{eqnarray}
\frac{1}{2}\int d\cos\theta~|\phi(\epsilon,\cos\theta)|^2_{\rm classical}&=\frac{q_0}{q}=\sqrt{1-\epsilon}
\end{eqnarray}
for both the attractive and repulsive cases.

When applying classical approximations for the wave function in Koonin's formula, one should be mindful of the divergences shown in Fig. \ref{fig:classical}. They are integrable, and the expressions remain tenable in a Monte-Carlo sampling procedure given sufficient sampling. However, the divergences do bring along a good deal of noise, even for the small values of $\epsilon$ used in the studies here. In the cases studied here, where the classical expressions are only applied for $q>500$ MeV/$c$, typical values of $\epsilon$ are $\sim 0.001$.

\begin{acknowledgments}
This work was supported by the Department of Energy Office of Science through grant number DE-FG02-03ER41259 and by the National Science Foundation's CSSI Program under Award Number OAC-2004601 (BAND Collaboration). The authors are grateful for the assistance with fitting routines provided by Ozge Surer.
\end{acknowledgments}

\end{document}